\documentclass[a4paper,UKenglish,cleveref, autoref]{lipics-v2019}
\usepackage{complexity}
\usepackage{upgreek}
\usepackage[ruled,vlined,titlenotnumbered]{algorithm2e}
\usepackage{comment}
\usepackage{tikz}
\usetikzlibrary{backgrounds}
\usetikzlibrary{calc}

\usepackage{tcolorbox}

\newcommand{\pw}{\mathrm{pw}}
\newcommand{\tw}{\mathrm{tw}}

\theoremstyle{plain}
\newtheorem*{question*}{Question}

\title{Finding and counting permutations via CSPs} %TODO Please add

\titlerunning{Finding and counting permutations via CSPs}%optional, please use if title is longer than one line

\author{Benjamin Aram Berendsohn}{Institut f\"ur Informatik, Freie Universit\"at Berlin}{beab@zedat.fu-berlin.de}{}{}%TODO mandatory, please use full name; only 1 author per \author macro; first two parameters are mandatory, other parameters can be empty. Please provide at least the name of the affiliation and the country. The full address is optional

\author{L\'aszl\'o Kozma}{Institut f\"ur Informatik, Freie Universit\"at Berlin}{laszlo.kozma@fu-berlin.de}{}{}%TODO mandatory, please use full name; only 1 author per \author macro; first two parameters are mandatory, other parameters can be empty. Please provide at least the name of the affiliation and the country. The full address is optional

\author{D\'aniel Marx}{Institute for Computer Science and Control, Hungarian Academy of Sciences (MTA SZTAKI)}{dmarx@cs.bme.hu}{}{}

\authorrunning{B.~Berendsohn, L.~Kozma, D.~Marx}%TODO mandatory. First: Use abbreviated first/middle names. Second (only in severe cases): Use first author plus 'et al.'

\Copyright{Benjamin Aram Berendsohn, L\'aszl\'o Kozma, D\'aniel Marx}%TODO mandatory, please use full first names. LIPIcs license is "CC-BY";  http://creativecommons.org/licenses/by/3.0/


%TODO mandatory; please add comma-separated list of keywords

\category{}%optional, e.g. invited paper

\relatedversion{}%optional, e.g. full version hosted on arXiv, HAL, or other respository/website
\supplement{}%optional, e.g. related research data, source code, ... hosted on a repository like zenodo, figshare, GitHub, ...

\acknowledgements{}%optional

\nolinenumbers %uncomment to disable line numbering

\hideLIPIcs  %uncomment to remove references to LIPIcs series (logo, DOI, ...), e.g. when preparing a pre-final version to be uploaded to arXiv or another public repository

\EventEditors{John Q. Open and Joan R. Access}
\EventNoEds{2}
\EventLongTitle{42nd Conference on Very Important Topics (CVIT 2016)}
\EventShortTitle{CVIT 2016}
\EventAcronym{CVIT}
\EventYear{2016}
\EventDate{December 24--27, 2016}
\EventLocation{Little Whinging, United Kingdom}
\EventLogo{}
\SeriesVolume{42}
\ArticleNo{23}
\begin{document}

\maketitle

%TODO mandatory: add short abstract of the document
\begin{abstract}
Permutation patterns and pattern avoidance have been intensively studied in combinatorics and computer science, going back at least to the seminal work of Knuth on stack-sorting (1968). Perhaps the most natural algorithmic question in this area is deciding whether a given permutation of length $n$ contains a given pattern of length $k$. 

In this work we give two new algorithms for this well-studied problem, one whose running time is $n^{k/4 + o(k)}$, and a polynomial-space algorithm whose running time is the better of $O(1.6181^n)$ and $O(n^{k/2 + 1})$. These results improve the earlier best bounds of $n^{0.47k + o(k)}$ and $O(1.79^n)$ due to Ahal and Rabinovich (2000) resp.\ Bruner and Lackner (2012) and are the fastest algorithms for the problem when $k \in \Omega(\log{n})$. 
We show that both our new algorithms and the previous exponential-time algorithms in the literature can be viewed through the unifying lens of \emph{constraint-satisfaction}. 

Our algorithms can also \emph{count}, within the same running time, the number of occurrences of a pattern.  We show that this result is close to optimal: solving the counting problem in time $f(k) \cdot n^{o(k/\log{k})}$ would contradict the \emph{exponential-time hypothesis} (ETH). For some special classes of patterns we obtain improved running times. We further prove that \emph{$3$-increasing} and \emph{$3$-decreasing} permutations can, in some sense, \emph{embed} arbitrary permutations of almost linear length, which indicates that an algorithm with sub-exponential running time is unlikely, even for patterns from these restricted classes. 

\end{abstract}

\newpage

\section{Introduction}
\label{sec:intro}

Let $[n] = \{1,\dots,n\}$. Given two permutations $\tau:[n] \rightarrow [n]$, % = (t_1, \dots, t_n)$ 
and $\pi:[k] \rightarrow [k]$,
%$\pi = (\pi_1, \dots, \pi_k)$, 
we say that $\tau$ \emph{contains} $\pi$, if there are indices $1 \leq i_1 < \cdots < i_k \leq n$ such that $\tau({i_j}) < \tau({i_\ell})$ if and only if $\pi({j}) < \pi(\ell)$, for all $1 \leq j,\ell \leq k$. In other words, $\tau$ contains $\pi$, if the sequence $(\tau(1), \dots, \tau(n))$ has a (possibly non-contiguous) subsequence with the same ordering as $(\pi(1),\dots,\pi(k))$, otherwise $\tau$ \emph{avoids} $\pi$. For example, $\tau = (1,5,4,6,3,7,8,2)$ contains $(2,3,1)$, because its subsequence $(5,6,3)$ has the same ordering as $(2,3,1)$; on the other hand, $\tau$ avoids $(3,1,2)$. 

Knuth showed in 1968~\cite[§\,2.2.1]{knuth68}, that permutations sortable by a single stack are exactly those that avoid $(2,3,1)$. Sorting by restricted devices has remained an active research topic~\cite{TarjanSorting, Pratt, Rosenstiehl, bona2003survey,AlbertB15, Pantone}, but permutation pattern avoidance has also taken on a life of its own (especially after the influential work of Simion and Schmidt~\cite{Simion}), becoming an important subfield of combinatorics. For more background on permutation patterns and pattern avoidance we refer to the extensive survey~\cite{vatter2014} and relevant textbooks~\cite{bona1, bona2, kitaev}. 

Perhaps the most important enumerative result related to permutation patterns is the theorem of Marcus and Tardos~\cite{MT} from 2004, stating that the number of length-$n$ permutations that avoid a fixed pattern $\pi$ is bounded by ${c(\pi)}^n$, where $c(\pi)$ is a quantity independent of $n$. (This was conjectured by Stanley and Wilf in the late 1980s.)  

A fundamental algorithmic problem in this context is \emph{Permutation Pattern Matching} (PPM): 
Given a length-$n$ permutation $\tau$ (``text'') and a length-$k$ permutation $\pi$ (``pattern''), decide whether $\tau$ contains $\pi$. 

Solving PPM is a bottleneck in experimental work on permutation patterns~\cite{Albert_algo}. The problem and its variants also arise in practical applications, e.g.\ in computational biology~\cite[§\,2.4]{kitaev} and time-series analysis~\cite{timeseries1, timeseries2, timeseries3}. Unfortunately PPM is, in general, $\NP$-complete, as shown by Bose, Buss, and Lubiw~\cite{BBL} in 1998. For small (e.g.\ constant-sized) patterns, the problem is solvable in polynomial (in fact, linear) time, as shown by Guillemot and Marx~\cite{GM} in 2013. Their algorithm has running time $n \cdot 2^{O(k^2 \log{k})}$, establishing the \emph{fixed-parameter tractability} of the PPM problem in terms of the pattern length. The algorithm builds upon the Marcus-Tardos proof of the Stanley-Wilf conjecture and introduces a novel decomposition of permutations. Subsequently, Fox~\cite{JFox} refined the Marcus-Tardos result, thereby removing a factor $\log{k}$ from the exponent of the Guillemot-Marx bound. (Due to the large constants involved, it is however, not clear whether the algorithm can be efficient in practice.)  

For longer patterns, e.g.\ for $k \in \Omega{(\log{n})}$, the complexity of the PPM problem is less understood. 
An obvious algorithm with running time $O(n^{k+1})$ is to enumerate all ${n \choose k}$ length-$k$ subsequences of $\tau$, checking whether any of them has the same ordering as $\pi$. 
The first result to break this ``triviality barrier'' was the $O(n^{2k/3+1})$-time algorithm of Albert, Aldred, Atkinson, and Holton~\cite{Albert_algo}. Shortly thereafter, Ahal and Rabinovich~\cite{Ahal} obtained the running time $n^{0.47k+o(k)}$. 

The two algorithms are based on a similar dynamic programming approach: they embed the entries of the pattern $\pi$ one-by-one into the text $\tau$, while observing the restrictions imposed by the current partial embedding. The order of embedding (implicitly) defines a \emph{path-decomposition} of a certain graph derived from the pattern $\pi$, called the \emph{incidence graph}. The running time obtainable in this framework is of the form $O(n^{\pw(\pi)+1})$, where $\pw(\pi)$ is the \emph{pathwidth} of the incidence graph of $\pi$.

Ahal and Rabinovich also describe a different, \emph{tree-based} dynamic programming algorithm that solves PPM in time $O(n^{2\cdot \tw(\pi) + 1})$, where $\tw(\pi)$ is the \emph{treewidth} of the incidence graph of $\pi$. Using known bounds on the treewidth, however, this running time does not improve the previous one. 

Our first result is based on the observation that PPM can be formulated as a \emph{constraint satisfaction problem} (CSP) with binary constraints. In this view, the path-based dynamic programming of previous works has a natural interpretation not observed earlier: it amounts to solving the CSP instance by \emph{Seidel's invasion algorithm}, a popular heuristic~\cite{SeidelCSP},\cite[§\,9.3]{CSPBOOK}.

It is well-known that binary CSP instances can be solved in time $O(n^{t+1})$, where $n$ is the \emph{domain size}, and $t$ is the \emph{treewidth} of the \emph{constraint graph}~\cite{treewidthCSP1, treewidthCSP2}. In our reduction, the domain size is the length $n$ of the text $\tau$, and the constraint graph is the incidence graph of the pattern $\pi$; we thus obtain a running time of $O(n^{\tw(\pi) + 1})$, improving upon the earlier $O(n^{2\cdot \tw(\pi) + 1})$. Second, making use of a bound known for low-degree graphs~\cite{FominGaspers}, we prove that the treewidth of the incidence graph of $\pi$ is at most $k/3 + o(k)$. The final improvement from $k/3$ to $k/4$ is achieved via a technique inspired by recent work of Cygan, Kowalik, and Soca\l{}a~\cite{Cygan} on the $k$-OPT heuristic for the \emph{traveling salesman problem} (TSP). 

In summary, we obtain the following result, proved in §\,\ref{sec:algcsp}.

\begin{theorem}\label{thm1}
Permutation Pattern Matching can be solved in time $n^{k/4 + o(k)}$.
\end{theorem}

Expressed in terms of $n$ only, none of the mentioned running times improve, in the worst case, upon the trivial $2^n$; consider the case of a pattern of length $k \geq n/\log{n}$. The first improvement in this parameter range was obtained by Bruner and Lackner~\cite{BL}; their algorithm runs in time $O(1.79^n)$.   

The algorithm of Bruner and Lackner works by decomposing both the text and the pattern into \emph{alternating runs} (consecutive sequences of increasing or decreasing elements), and using this decomposition to restrict the space of admissible matchings. The exponent in the running time is, in fact, the \emph{number of runs} of $T$, which can be as large as $n$. The approach is compelling and intuitive, the details, however, are intricate (the description of the algorithm and its analysis in~\cite{BL} take over 24 pages).

Our second result improves this running time to $O(1.618^n)$, with an exceedingly simple approach. A different analysis of our algorithm yields the bound $O(n^{k/2+1})$, i.e.\ slightly above the Ahal-Rabinovich bound~\cite{Ahal}, but with polynomial space. The latter bound also matches an earlier result of Guillemot and Marx~\cite[§\,7]{GM}, obtained via involved techniques.

\begin{theorem}\label{thm2}
  Permutation Pattern Matching can be solved using polynomial space, in time $O(1.6181^n)$ or $O(n^{k/2 + 1})$. 
\end{theorem}

At the heart of this algorithm is the following observation: if all \emph{even-index} entries of the pattern $\pi$ are matched to entries of the text $\tau$, then verifying whether the remaining \emph{odd-index} entries of $\pi$ can be correctly matched takes only a linear-time sweep through both $\pi$ and $\tau$. This algorithm can be explained very simply in the CSP framework: after substituting a value to every  \emph{even-index} variable, the graph of the remaining constraints is a union of paths, and hence can be handled very easily.

\subparagraph*{Counting patterns.}
We also consider the closely related problem of \emph{counting} the number of occurrences of $\pi$ in $\tau$, i.e.\ finding the number of subsequences of $\tau$ that have the same ordering as $\pi$. Easy modifications of our algorithms solve this problem within the bounds of Theorems~\ref{thm1} and \ref{thm2}.

\begin{theorem}\label{thm3}
  The number of solutions for Permutation Pattern Matching can be computed
  \begin{itemize}
  \item in time $n^{k/4+o(k)}$,
  \item in time $O(n^{k/2+2})$ and polynomial space, and
  \item in time $O(1.6181^n)$ and polynomial space.
  \end{itemize}
\end{theorem}

Note that the FPT algorithm of Guillemot and Marx \cite{GM} cannot be adapted for the counting version. In fact, we argue (§\,\ref{sec:hardness}) that a running time of the form $n^{O(k)}$ is almost best possible and a significant improvement in running time for the counting problem is unlikely.

\begin{theorem}\label{thmhard}
Assuming the exponential-time hypothesis (ETH), there is no algorithm that counts the number of occurrences of $\pi$ in $\tau$ in time $f(k) \cdot n^{o(k/\log{k})}$, for any function $f$.
\end{theorem}

\subparagraph*{Special patterns.}
It is possible that PPM is easier if the pattern $\pi$ comes from some restricted family of permutations, e.g.\ if it avoids some smaller fixed pattern $\sigma$. Several such examples have been studied in the literature, and recently Jel\'{i}nek and Kyn\u{c}l~\cite{hardness321} obtained the following characterization: PPM is polynomial-time solvable for $\sigma$-avoiding patterns $\pi$, if $\sigma$ is one of $(1)$, $(1,2)$, $(1,3,2)$, $(2,1,3)$ or their reverses, and $\NP$-complete for all other $\sigma$. 
(All tractable cases are such that $\pi$ is a \emph{separable} permutation~\cite{BBL, Ibarra, Saxena, Albert_algo}.)

In particular, Jel\'{i}nek and Kyn\u{c}l show that PPM is $\NP$-complete even if $\pi$ avoids $(1,2,3)$ or $(3,2,1)$, but polynomial-time solvable for any proper subclass of these families. For $(1,2,3)$-avoiding and $(3,2,1)$-avoiding patterns, it is known however, that PPM can be solved in time $n^{O(\sqrt{k})}$, i.e.\ faster than the general case~(Guillemot and Vialette \cite{321}).

These results motivate the following general and natural question. 

\begin{question*}
What makes a permutation pattern easier to find than others?
\end{question*}

A permutation is \emph{$t$-monotone}, if it can be obtained by interleaving $t$ monotone sequences. When all $t$ sequences are increasing (resp.\ decreasing), we call the resulting permutation $t$-increasing (resp.\ $t$-decreasing). It is well-known that $t$-increasing (resp.\ $t$-decreasing) permutations are exactly those that avoid $(t+1, \dots, 1)$, resp.\ $(1, \dots, t+1)$, see e.g.\ \cite{Aldous}.

We prove that if $\pi$ is $2$-monotone, then the running time of the algorithm of Theorem~\ref{thm1} is $n^{O(\sqrt{k})}$. This result follows from bounding the treewidth of the incidence graph of $\pi$, by observing that this graph is \emph{almost planar}. For $2$-increasing or $2$-decreasing patterns we thus match the bound of Guillemot and Vialette by a significantly simpler argument. (In these special cases the incidence graph is, in fact, planar.)

\emph{Jordan-permutations} are a natural family of geometrically-defined permutations with applications in computational geometry~\cite{rosenstiehl:hal-00259765}. They were studied by Hoffmann, Mehlhorn, Rosenstiehl, and Tarjan~\cite{hoffmann1986sorting}, who showed that they can be sorted with a linear number of comparisons (see also~\cite{meanders} for related enumerative results). A Jordan permutation is generated by the intersection-pattern of two simple curves in the plane: label the intersection points between the curves in increasing order along the first curve, and read out the labels along the second curve; the obtained sequence is a Jordan-permutation (Figure~\ref{fig1}). %We show that if the pattern is a Jordan-permutation, then PPM can be solved in sub-exponential time. 
As the 
%We show that the 
incidence graph of the pattern $\pi$ is planar whenever $\pi$ is a Jordan-permutation, in this case too % (this is, in fact, an exact characterization), from which 
an $n^{O(\sqrt{k})}$ bound on the running time %of the algorithm of Theorem~\ref{thm1} 
follows. %when $\pi$ is a Jordan-permutation.

\begin{theorem}\label{thmapp}
The treewidth of the incidence graph of $\pi$ is $O(\sqrt{k})$, \\ 
(i) if $\pi$ is $2$-monotone, or~~(ii) if $\pi$ is a Jordan-permutation.
\end{theorem}

We show that both $2$-monotone (and even $2$-increasing or $2$-decreasing) and Jordan-permutations of length $O(k)$ may contain grids of size $\sqrt{k} \times\sqrt{k}$ in their incidence-graphs, both statements of Theorem~\ref{thmapp} are therefore tight, via known lower bounds on the treewidth of grids~\cite{pathwidth}. 
 
In light of these results, one may try to obtain further treewidth-bounds for families of patterns, in order to solve PPM in sub-exponential time. In this direction we show a (somewhat surprising) negative result. 

\begin{theorem}\label{thmembed}
There are $3$-increasing permutations of length $k$ whose incidence graph has treewidth $\Omega(k/\log{k})$.
\end{theorem}

The same bound applies, by symmetry, to $3$-decreasing permutations. The result is obtained by embedding the incidence graph of an \emph{arbitrary} permutation of length $O(k/\log{k})$ as a \emph{minor} of the incidence graph of a $3$-increasing permutation of length $k$.

Theorems~\ref{thmapp} and~\ref{thmembed} (proved in §\,\ref{sec:spec}) lead to an almost complete characterization of the treewidth of $\sigma$-avoiding patterns. 
%%%For $|\sigma| \leq 3$, tight bounds on $\tw(\pi)$ follow from known results and the preceding discussion. 
By the Erd\H{o}s-Szekeres theorem~\cite{ESz} every $k$-permutation contains a monotone pattern of length $\lceil \sqrt{k} \rceil$. Thus, for all permutations $\sigma$ of length at least $10$, the class of $\sigma$-avoiding permutations contains all $3$-increasing or all $3$-decreasing permutations, hence by Theorem~\ref{thmembed} there exist $\sigma$-avoiding patterns $\pi$ with $\tw(\pi) \in \Omega{({k}/\log{k})}$. Addressing a few additional small cases by similar arguments (details given in the forthcoming thesis of the first author), the threshold $10$ can be further reduced. %in the worst-case $|\sigma|$-avoiding patterns $\pi$ for
We remark that no algorithm is known to solve PPM in time $n^{o(\tw(\pi))}$; see the discussion in~\cite{Ahal, hardness321}.

With a weaker bound we obtain a full characterisation that strengthens the dichotomy-result of Jel\'{i}nek and Kyn\u{c}l: in the worst case, the 
%%
%%Addressing one additional case (showing that $(2,1,4,3)$- and $(3,1,4,2)$-avoiding permutations can have treewidth $\Omega(\sqrt{n})$, we obtain that, in the worst case, the 
only $\sigma$-avoiding patterns $\pi$ for which $\tw{(\pi)} \in o{(\sqrt{k})}$ are those for which PPM is known to be polynomial-time solvable (Appendix~\ref{appq}). 

\subparagraph*{Further related work.} 

The complexity of the PPM problem has also been studied under the stronger restriction that the text $\tau$ is pattern-avoiding. The problem is polynomial-time solvable if $\tau$ is monotone~\cite{Chang} or $2$-monotone~\cite{vialette, 321, albert321, Albert_algo, 231}, but NP-hard if $\tau$ is $3$-monotone~\cite{hardness321}. A broader characterization is missing. 

Only \emph{classical patterns} are considered in this paper; variants in the literature include \emph{vincular}, \emph{bivincular}, \emph{consecutive}, and \emph{mesh} patterns; we refer to~\cite{BLcomp} for a survey of related computational questions.

Newman et al.\ \cite{Newman} study pattern matching in a \emph{property-testing} framework (aiming to distinguish pattern-avoiding sequences from those that contain \emph{many copies} of the pattern). In this setting, the focus is on the \emph{query complexity} of different approaches, and sampling techniques are often used; see also~\cite{BenEliezerC, foxPT}.

A different line of work investigates whether standard algorithmic problems on permutations (e.g.\ sorting, selection) become easier if the input can be assumed to be pattern-avoiding~\cite{Arthur, FOCS15}.

\section{Preliminaries}\label{sec:prel}

A length-$n$ permutation $\sigma$ is a bijective function $\sigma: [n] \rightarrow [n]$, alternatively viewed as the sequence $(\sigma(1), \dots, \sigma(n))$. Given a length-$n$ permutation $\sigma$, we denote as $S_\sigma = \{(i,\sigma(i)) \mid 1 \leq i \leq n\}$ the \emph{set of points} corresponding to permutation $\sigma$. 

For a point $p \in S_\sigma$ we denote its first entry as $p.x$, and its second entry as $p.y$, referring to these values as the \emph{index}, respectively, the \emph{value} of $p$. Observe that for every $i \in [n]$, we have $|\{p\in S_\sigma \mid p.x = i\}| = |\{p\in S_\sigma \mid p.y = i\}| = 1$.

We define four neighbors of a point $(x,y) \in S_\sigma$ as follows.
\begin{eqnarray*}
N^R((x,y)) & = & (x+1,~ \sigma(x+1)), \\
N^L((x,y)) & = & (x-1,~ \sigma(x-1)), \\
N^U((x,y)) & = & (\sigma^{-1}(y+1),~ y+1), \\
N^D((x,y)) & = & (\sigma^{-1}(y-1),~ y-1). 
\end{eqnarray*}

The superscripts $R$, $L$, $U$, $D$ are meant to evoke the directions \emph{right}, \emph{left}, \emph{up}, \emph{down}, when plotting $S_\sigma$ in the plane. Some neighbors of a point may coincide. When some index is out of bounds, we let the offending neighbor be a ``virtual point'' as follows: $N^R(n,i) = N^U(i,n) = (\infty,\infty)$, and $N^L(1,i) = N^D(i,1) = (0,0)$, for all $i \in [n]$. The virtual points are not contained in $S_\sigma$, we only define them to simplify some of the statements.

\begin{figure*}[h]
  \centering
  \includegraphics[width=10cm]{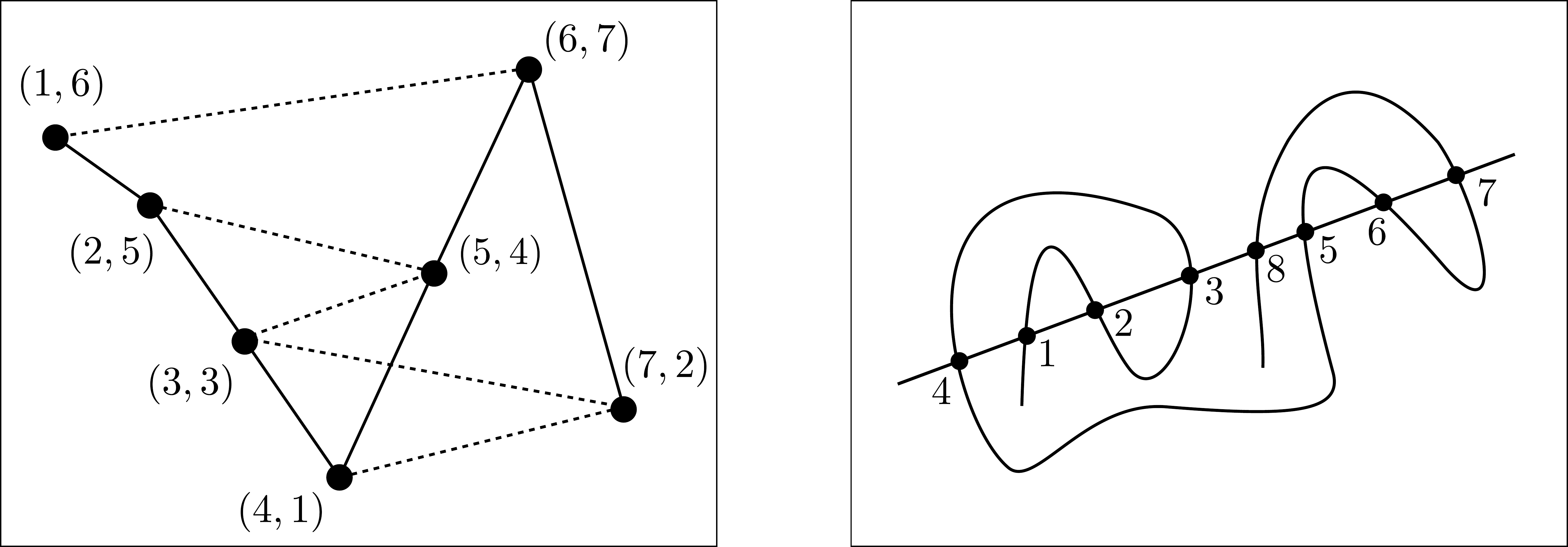}
  \caption{(\emph{left}) Permutation $\pi = (6,5,3,1,4,7,2)$ and its incidence graph $G_\pi$. Solid lines indicate neighbors by index, dashed lines indicate neighbors by value (lines may overlap). Indices plotted on $x$-coordinate, values plotted on $y$-coordinate. (\emph{right}) Jordan-permutation $(4,1,2,3,8,5,6,7)$\label{fig1}}
\end{figure*}

The \emph{incidence graph} of a permutation $\sigma$ is $G_\sigma = (S_\sigma,E_\sigma)$, where $$E_\sigma = \left\{\left(p,N^{\upalpha}(p)\right) \mid \upalpha \in \{R,L,U,D\} \mbox{~and~~} p, N^{\upalpha}(p) \in S_\sigma\right\}.$$ In words, each point is connected to its (at most) four neighbors: its successor and predecessor by index, and its successor and predecessor by value. It is easy to see that $G_\sigma$ is a union of two Hamiltonian paths on the same set of vertices and that this is an exact characterization of permutation incidence-graphs. (See Figure~\ref{fig1} for an illustration.)

Throughout the paper we consider a text permutation $\tau: [n] \rightarrow [n]$, and a pattern permutation $\pi : [k] \rightarrow [k]$, where $n \geq k$. We give an alternative definition of the Permutation Pattern Matching (PPM) problem in terms of embedding $S_\pi$ into $S_\tau$. 

Consider a function $f:S_\pi \rightarrow S_\tau$. We say that $f$ is a \emph{valid embedding} of $S_\pi$ into $S_\tau$ if for all $p \in S_\pi$ the following hold:
\begin{eqnarray}
f(N^L(p)).x & < ~~f(p).x  & < ~~f(N^R(p)).x, \mbox{~~and}\\
f(N^D(p)).y & < ~~f(p).y & < ~~f(N^U(p)).y, 
\end{eqnarray}
whenever the corresponding neighbor $N^{\upalpha}(p)$ is also in $S_\pi$, i.e.\ not a virtual point. In words, valid embeddings preserve the relative positions of neighbors in the incidence graph.

\begin{lemma}[Proof in Appendix~\ref{appa}]\label{eqlem}
Permutation $\tau$ contains permutation $\pi$ if and only if there exists a valid embedding $f:S_\pi \rightarrow S_\tau$.
\end{lemma}

For sets $A \subseteq B \subseteq S_\pi$ and functions $g : A \rightarrow S_\tau$ and $f: B \rightarrow S_\tau$ we say that $g$ is the \emph{restriction} of $f$ to $A$, denoted $g = f |_{A}$, if  $g(i) = f(i)$ for all $i \in A$. In this case, we also say that $f$ is the \emph{extension} of $g$ to $B$. 
Restrictions of valid embeddings will be called \emph{partial embeddings}. We observe that if $f : B \rightarrow S_\tau$ is a partial embedding, then it satisfies conditions (1) and (2) with respect to all edges in the induced graph $G_\pi[B]$, i.e.\ the corresponding inequality holds whenever $p, N^\upalpha(p) \in B$. 

\section{Pattern matching as constraint satisfaction}\label{sec:algcsp}

Readers familiar with the terminology of CSPs should immediately recognize that the definition of valid embedding and Lemma~\ref{eqlem} allow us to formulate PPM as a CSP instance with binary constraints. Then known techniques can be applied to solve the problem. A (somewhat different) connection of PPM to CSPs was previously observed by Guillemot and Marx~\cite{GM}. We first review briefly the CSP problem, referring to~\cite{CSPBOOK, SeidelCSP, surveyCSP} for more background.

A \emph{binary CSP} instance is a triplet $(V,D,C)$, where $V$ is a set of variables, $D$ is a set of admissible values (the \emph{domain}), and $C$ is a set of constraints $C = \{c_1, \dots, c_m\}$, where each constraint $c_i$ is of the form $((x,y),R)$, where $x,y \in V$, and $R \subseteq D^2$ is a binary relation.

A solution of the CSP instance is a function $f: V \rightarrow D$ (i.e.\ an assignment of admissible values to the variables), such that for each constraint $c_i = ((x_i,y_i),R_i)$, the pair of assigned values $(f(x_i),f(y_i))$ is contained in $R_i$.

The reduction from PPM to CSP is straightforward. Given a PPM instance with text $\tau$ and pattern $\pi$, of lengths $n$ and $k$ respectively, let $V = \{x_1, \dots, x_k \}$, and $D = \{1, \dots, n\}$. 
The fact that variable $x_i$ takes value $j$ signifies that $\pi(i)$ is matched (embedded) to $\tau(j)$. For the embedding to be valid, by Lemma~\ref{eqlem}, the relative ordering of entries must be respected, in accordance with conditions (1) and (2). These conditions can readily be described by binary relations for all pairs of variables whose corresponding entries are neighbors in the incidence graph $G_\pi$.

More precisely, for $p, N^{\upalpha}(p) \in S_\pi$, for $\upalpha \in \{R,L,U,D\}$, we add constraints of the form $((x_i,x_j),R)$, where $i = p.x$, $j = {N^{\upalpha}(p).x}$ and $R$ contains those pairs $(a,b) \in [n]^2$, for which the relative position of $(a,\tau(a))$ and $(b,\tau(b))$ matches the relative position of $p$ and $N^{\upalpha}(p)$.

The \emph{constraint graph} of the binary CSP instance (also known as \emph{primal graph} or \emph{Gaifman graph}) is a graph whose vertices are the variables $V$ and whose edges connect all pairs of variables that occur together in a constraint. Observe that for instances obtained via our reduction, the constraint graph is exactly the incidence graph $G_\pi$. We make use of the following well-known result.

\begin{lemma}[\cite{treewidthCSP1, treewidthCSP2}] \label{lemtr1}
A binary CSP instance $(V,D,C)$ can be solved in time $O(|D|^{t+1})$ where $t$ is the {treewidth} of the constraint graph.
\end{lemma}

As discussed in §\,\ref{sec:prel}, the incidence graph $G_\pi$ consists of two Hamiltonian-paths. Accordingly, its vertices have degree at most $4$, and the following structural result is applicable.

\begin{lemma}[\cite{FominGaspers, FominHoie}] \label{lemtr2}
If $G$ is an order-$k$ graph with vertices of degree at most $4$, then the pathwidth (and consequently, the treewidth) of $G$ is at most $k/3 + o(k)$. A corresponding tree-(path-)decomposition can be found in polynomial time.
\end{lemma}

\subparagraph*{Algorithms.} Our first algorithm amounts to reducing the PPM instance to a binary CSP instance, and using the algorithm of Lemma~\ref{lemtr1} with a tree-decomposition obtained via Lemma~\ref{lemtr2}. To reach the bound given in Theorem~\ref{thm1}, it remains to improve the $k/3$ term in the exponent to $k/4$. We achieve this with a recent technique of Cygan et al.~\cite{Cygan}, developed in the context of the $k$-OPT heuristic for TSP. 

In our setting, the technique works as follows. We split $[n]$ into $n^{1/4}$ contiguous intervals of equal widths, $n^{3/4}$ each. (For simplicity, we ignore issues of rounding and divisibility.) The intervals induce vertical \emph{strips} in the text $\tau$. For each pattern-index $i \in [k]$ we \emph{guess} the vertical strip of $\tau$ into which $i$ is mapped in the sought-for embedding of $\pi$ into $\tau$. It is sufficient to do this for a subset of the entries in $\pi$, namely those that become the \emph{leftmost} in their respective strips in $\tau$. Let $X \subseteq [k]$ be the set of indices of such entries in $\pi$. 

Guessing $X$ and the strips of $\tau$ into which entries of $X$ are mapped increases the running time by a factor of $\sum_{X \subseteq [k]}{n^{1/4} \choose |X|} \leq \sum_{X \subseteq [k]} {n^{|X|/4}}$. Assuming that we guessed correctly, the problem simplifies. First, each pattern-entry can now be embedded into at most $n^{3/4}$ possible locations, hence the domain of each variable will be of size at most $n^{3/4}$. Second, the horizontal constraints that go across strip-boundaries can now be removed as they are implicitly enforced by the distribution of entries into strips (the $L$-constraint of every $X$-entry is removed). We have thus reduced the number of edges in the constraint-graph by $|X|-1$ and can use a stronger upper bound of $(k- |X|)/3 + o(k)$ on the treewidth (see e.g.\ \cite{FominGaspers, Cygan}). 
The overall running time becomes $$\sum_{X \subseteq [k]} n^{\frac{|X|}{4}} \cdot n^{\frac{3}{4} \cdot (\frac{k}{3} - \frac{|X|}{3}) + o(k)} = 2^k \cdot n^{k/4 + o(k)} = n^{k/4 + o(k)}.$$

We remark that our use of this technique is essentially the same as in Cygan et al.~\cite{Cygan}, but the CSP-formalism makes its application more transparent. We suspect that further classes of CSPs could be handled with a similar approach. 

\subparagraph*{The even-odd method.} The algorithm for Theorem~\ref{thm2} can be obtained as follows.
Let $(Q^E, Q^O)$ be the partition of $S_\pi$ into points with even and odd indices. Formally, $Q^{E} = \{(2k, \pi(2k)) \mid 1 \leq k \leq \lfloor k/2 \rfloor \}$, and $Q^{O} = \{(2k-1,\pi(2k-1)) \mid 1 \leq k \leq \lceil k/2 \rceil \}$. 
Construct the CSP instance corresponding to the problem as above. A solution is now found by trying first every possible combination of values for the variables representing $Q^E$. Clearly, there are $n^{|Q^E|}= n^{\lfloor k/2 \rfloor}$ possible combinations. If the value of a variable $x_i$ is fixed to $a\in [n]$, then $x_i$ is removed from the problem and every neighbor of $x_i$ is restricted by a new unary constraint  in an appropriate way, i.e.\ if there is a constraint $((x_i,x_j),R)$, then $x_j$ should be restricted to values $b$ for which $(a,b)\in R$.

How does the constraint graph look like if we remove every variable (and its incident edges) corresponding to $Q^E$? It is easy to see that this destroys every constraint corresponding to L-R neighbors and all the remaining binary constraints represent U-D neighbors. As these constraints form a Hamiltonian path, the remaining constraint graph consists of a union of disjoint paths. Such graphs have treewidth 1, hence the resulting CSP instance can be solved efficiently using Lemma~\ref{lemtr1}, resulting in the running time $O(n^{k/2+2})$. A more careful argument improves this bound to $O(n^{k/2+1})$; we describe the details in Appendix~\ref{appb}.

We can refine the analysis, noting that when we are assigning values $a_2<a_4<a_6<\dots$ to the variables $x_2$, $x_4$, $x_6$, $\dots$ representing $Q^E$, then we need to consider only increasing sequences where there is a gap of at least one between each successive entry (e.g.\ $a_{4}>a_{2}+1$) to allow a value for the odd-indexed variables. The number of such subsequences is ${{n-\lceil k/2 \rceil} \choose {\lfloor k/2 \rfloor}}$: consider a sequence with a minimum required gap of one between consecutive entries, then distribute the remaining total gap of $n-k$ among the $ \lfloor k/2 \rfloor + 1 $ slots. 
As $\max_k{{n-k} \choose k} =  O(1.6181^n)$, see e.g.\ \cite{fibo_seq1, fibo_seq2}, we obtain an upper bound of this form (independent of $k$) on the running time of the algorithm.

\subparagraph*{Counting solutions.} %We would like to point out that.
The algorithms described above can be made to work for the counting version of the problem. % with standard modifications.
 This has to be contrasted with the FPT algorithm of Guillemot and Marx~\cite{GM}, which cannot be adapted for the counting version: a crucial step in that algorithm is to say that if the text is sufficiently complicated, then it contains every pattern of length $k$, hence we can stop. Indeed, as we show in §\,\ref{sec:hardness}, we cannot expect an FPT algorithm for the counting problem.

To solve the counting problem, we modify the dynamic programming algorithm behind Lemma~\ref{lemtr1} in a straightforward way. Even if not stated in exactly the following form, results of this type are implicitly used in the counting literature.
\begin{lemma}\label{lemcount}
The number of solutions of a binary CSP instance $(V,D,C)$ can be computed in time $O(|D|^{t+1})$ where $t$ is the {treewidth} of the constraint graph.
\end{lemma}
It is not difficult to see that by replacing the use of Lemma~\ref{lemtr1} with Lemma~\ref{lemcount} in the algorithms of Theorems~\ref{thm1} and \ref{thm2}, the counting algorithms stated in Theorem~\ref{thm3} follow.

\section{Special patterns}\label{sec:spec}

In this section we  prove Theorems~\ref{thmapp} and \ref{thmembed}. We define a \emph{$k$-track graph} $G = (V, E)$ to be the union of two Hamiltonian paths $H_1$ and $H_2$, where $V$ can be partitioned into \emph{sequences} $S_1, S_2, \dots, S_k$, the \emph{tracks} of $G$, such that both $H_1$ and $H_2$ visit the vertices of $S_i$ in the given order, for all $i \in [k]$. Observe that $k$-track graphs are exactly the incidence graphs of permutations that are either $k$-increasing or $k$-decreasing.

\subparagraph*{$2$-monotone patterns.} \label{sec2mon}

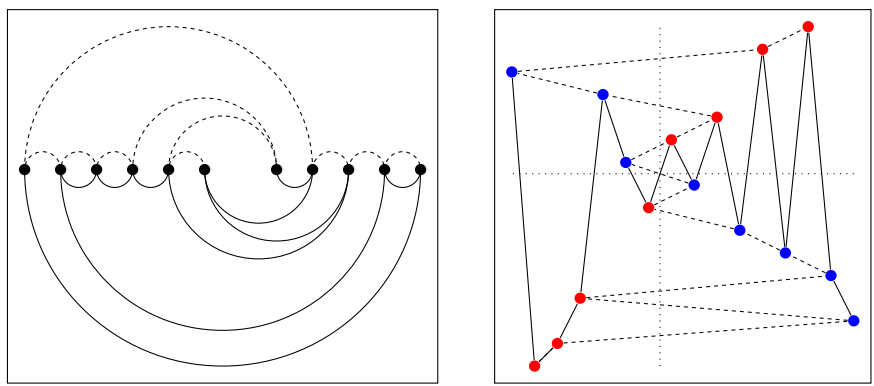
\begin{figure}[h]
	
	\centering
	\begin{tikzpicture}[
		scale={4.5/9.5},
		framed,
		baseline=(current bounding box.center),
		vertex/.style={circle, fill, inner sep=1.5pt},
		top_arc/.style={dash pattern={on 1.5pt off 1.5pt}},
		bottom_arc/.style={}
	]
	\node[vertex](x_1) at (1, 0) {};
\node[vertex](x_2) at (2, 0) {};
\node[vertex](x_3) at (3, 0) {};
\node[vertex](x_4) at (4, 0) {};
\node[vertex](x_5) at (5, 0) {};
\node[vertex](y_1) at (-1, 0) {};
\node[vertex](y_2) at (-2, 0) {};
\node[vertex](y_3) at (-3, 0) {};
\node[vertex](y_4) at (-4, 0) {};
\node[vertex](y_5) at (-5, 0) {};
\node[vertex](y_6) at (-6, 0) {};
\draw [top_arc] (-1,0) arc (0:180:0.5);
\draw [top_arc] (-2,0) arc (180:0:1.5);
\draw [top_arc] (1,0) arc (0:180:2.0);
\draw [top_arc] (-3,0) arc (0:180:0.5);
\draw [top_arc] (-4,0) arc (0:180:0.5);
\draw [top_arc] (-5,0) arc (0:180:0.5);
\draw [top_arc] (-6,0) arc (180:0:4.0);
\draw [top_arc] (2,0) arc (180:0:0.5);
\draw [top_arc] (3,0) arc (180:0:0.5);
\draw [top_arc] (4,0) arc (180:0:0.5);
\draw [bottom_arc] (1,0) arc (180:360:0.5);
\draw [bottom_arc] (2,0) arc (360:180:1.5);
\draw [bottom_arc] (-1,0) arc (180:360:2.0);
\draw [bottom_arc] (3,0) arc (360:180:2.5);
\draw [bottom_arc] (-2,0) arc (360:180:0.5);
\draw [bottom_arc] (-3,0) arc (360:180:0.5);
\draw [bottom_arc] (-4,0) arc (360:180:0.5);
\draw [bottom_arc] (-5,0) arc (180:360:4.5);
\draw [bottom_arc] (4,0) arc (180:360:0.5);
\draw [bottom_arc] (5,0) arc (360:180:5.5);
	% Invisible nodes to fix size
	\node[circle, inner sep=1.5pt] at (0,4) {};
	\node[circle, inner sep=1.5pt] at (0,-5.5) {};
	\end{tikzpicture}
	\hspace{5mm}
	\begin{tikzpicture}[
		scale={4.5/15},
		rotate={270},
		framed,
		baseline=(current bounding box.center),
		v1/.style={circle, fill, inner sep=1.5pt, red},
		v2/.style={circle, fill, inner sep=1.5pt, blue},
		p1/.style={dash pattern={on 1.5pt off 1.5pt}},
		p2/.style={}
	]
	\node[v1](1) at (0, 14) {};
\node[v1](2) at (1, 12) {};
\node[v2](a) at (2, 1) {};
\node[v2](b) at (3, 5) {};
\node[v1](3) at (4, 10) {};
\node[v1](4) at (5, 8) {};
\node[v2](c) at (6, 6) {};
\node[v2](d) at (7, 9) {};
\node[v1](5) at (8, 7) {};
\node[v2](e) at (9, 11) {};
\node[v2](f) at (10, 13) {};
\node[v2](g) at (11, 15) {};
\node[v1](6) at (12, 4) {};
\node[v2](h) at (13, 16) {};
\node[v1](7) at (14, 3) {};
\node[v1](8) at (15, 2) {};
\draw [p1] (2) -- (1);
\draw [p1] (a) -- (2);
\draw [p1] (b) -- (a);
\draw [p1] (3) -- (b);
\draw [p1] (4) -- (3);
\draw [p1] (c) -- (4);
\draw [p1] (d) -- (c);
\draw [p1] (5) -- (d);
\draw [p1] (e) -- (5);
\draw [p1] (f) -- (e);
\draw [p1] (g) -- (f);
\draw [p1] (6) -- (g);
\draw [p1] (h) -- (6);
\draw [p1] (7) -- (h);
\draw [p1] (8) -- (7);
\draw [p2] (8) -- (a);
\draw [p2] (7) -- (8);
\draw [p2] (6) -- (7);
\draw [p2] (b) -- (6);
\draw [p2] (c) -- (b);
\draw [p2] (5) -- (c);
\draw [p2] (4) -- (5);
\draw [p2] (d) -- (4);
\draw [p2] (3) -- (d);
\draw [p2] (e) -- (3);
\draw [p2] (2) -- (e);
\draw [p2] (f) -- (2);
\draw [p2] (1) -- (f);
\draw [p2] (g) -- (1);
\draw [p2] (h) -- (g);
\draw [dotted] (15,7.5) -- (0,7.5);
\draw [dotted] (6.5,16) -- (6.5,1);
	\end{tikzpicture}
	\caption{(\emph{left}) A planar drawing of a $2$-track graph, with one of the two Hamiltonian paths drawn with dashed arcs. Note that edges contained in both Hamiltonian paths are drawn twice for clarity.
		(\emph{right}) A drawing of the incidence graph of a $2$-monotone permutation. Red and blue dots indicate an increasing (resp.\ decreasing) subsequence.	\label{fig:planar_2mon}
}
\end{figure}

We prove Theorem~\ref{thmapp}(i). As a special case, we first look at patterns that are $2$-increasing. Let $G$ be a $2$-track graph. Arrange the vertices of the two tracks on a line $\ell$, the first track in reverse order, followed by the second track in sorted order. Any Hamiltonian path that respects the order of the two tracks can be drawn (without crossings) on one side of $\ell$. This means that the two Hamiltonian paths of $G$ can be drawn on different sides of $\ell$, and therefore $G$ is planar. See Figure~\ref{fig:planar_2mon} (left) for an example.

The treewidth of a $k$-vertex planar graph is known to be $O(\sqrt{k})$~\cite{pathwidth,DowneyFellows}. A corresponding path-decomposition can be obtained by a recursive use of planar separators. %A corresponding embedding order $\tau$ of $\pi$ can be built recursively, concatenating the sequences obtained on the different sides of the separator and the sequence obtained from the separator itself.
For the case of a pattern $\pi$ that consists of an increasing and a decreasing subsequence (i.e.\ $2$-monotone patterns), we show that the straight-line drawing of $G_\pi$ (with points $S_\pi$ as vertices) has at most one intersection. An $O(\sqrt{k})$ bound on the treewidth follows via known results~\cite{pwcross}.

Divide $S_\pi$ by one horizontal and one vertical line, such that each of the resulting four sectors contains a monotone sequence. More precisely, the top left and bottom right sectors contain decreasing subsequences, and the other two sectors contain increasing subsequences. Let $e = \{u,v\}$ be an edge of the horizontal Hamiltonian path such that $u.x = v.x - 1$, and let $f = \{s,t\}$ be an edge that intersects $e$, such that $s.x < t.x$. Edge $f$ must come from the vertical Hamiltonian path, i.e.\ $|s.y - t.y| = 1$. As $u$ and $v$ are horizontal neighbors, $s.x < u.x < v.x < t.x$ holds. Assume that $u.y < v.y$ (otherwise flip $G_\pi$ vertically before the argument, without affecting the graph structure).  
We claim that $u.y  < t.y < s.y < v.y$ must hold, as otherwise $\pi$ contains the pattern $(2,1,4,3)$ and cannot decompose into an increasing and a decreasing subsequence.
 
Thus $(s,u,v,t)$ must form the pattern $(3,1,4,2)$, and therefore $s$ and $t$ belong to the decreasing and $u$ and $v$ to the increasing subsequence. It is easy to see now that $s,u,v,t$ must be in pairwise distinct sectors, and $u$ ($s$, $t$, $v$) is the unique rightmost (bottommost, topmost, leftmost) point of the bottom left (top left, top right, bottom right) sector, and due to the monotonicity of all four sectors no more intersections can happen; see Figure~\ref{fig:planar_2mon} (right).   
%The second case is symmetric, with $s,u,v,t$ forming the pattern $(2,4,1,3)$.
This concludes the proof of Theorem~\ref{thmapp}(i).

\medskip

We show that the previous result is tight, by constructing a $2$-track graph $G = (V,E)$ with $n = 2 k^2$ vertices, for some even $k$, that contains a $k \times 2k$ grid graph. 

 Let $x_1, x_2, \dots, x_{k^2}$ and $y_1, y_2, \dots, y_{k^2}$ be the two tracks of $G$. We obtain $G$ as the union of the following two Hamiltonian paths:
\begin{eqnarray*}
	&& x_1, y_1, y_2, x_2, x_3, y_3, y_4, x_4, \dots, x_{k^2-1}, y_{k^2-1}, y_{k^2}, x_{k^2}\mbox{; ~~ and} \\
	&& x_1, x_2, \dots, x_k,\\
	&& y_1, x_{k+1}, x_{k+2}, y_2, y_3, x_{k+3}, x_{k+4}, y_4, \dots y_{k^2-k-1} x_{k^2-1}, x_{k^2}, y_{k^2-k}, \\
	&& y_{k^2-k+1}, y_{k^2-k+2}, \dots, y_{k^2}.
\end{eqnarray*}

The two Hamiltonian paths respect the order of the two tracks.

\begin{figure}[h]
	
	\centering
	\begin{tikzpicture}[
		scale={5/31},
		framed,
		baseline=(current bounding box.center),
		px/.style={fill, circle, inner sep=1.5pt},
		py/.style={fill, rectangle, inner sep=2pt},
		c0/.style=blue,
		c1/.style=red,
		c2/.style={black!50!green},
		c3/.style=brown]
		\node[px, c1](x_1) at (0, 1) {};
\node[py, c1](y_1) at (1, 5) {};
\node[py, c2](y_2) at (2, 8) {};
\node[px, c2](x_2) at (3, 2) {};
\node[px, c3](x_3) at (4, 3) {};
\node[py, c3](y_3) at (5, 9) {};
\node[py, c0](y_4) at (6, 12) {};
\node[px, c0](x_4) at (7, 4) {};
\node[px, c1](x_5) at (8, 6) {};
\node[py, c1](y_5) at (9, 13) {};
\node[py, c2](y_6) at (10, 16) {};
\node[px, c2](x_6) at (11, 7) {};
\node[px, c3](x_7) at (12, 10) {};
\node[py, c3](y_7) at (13, 17) {};
\node[py, c0](y_8) at (14, 20) {};
\node[px, c0](x_8) at (15, 11) {};
\node[px, c1](x_9) at (16, 14) {};
\node[py, c1](y_9) at (17, 21) {};
\node[py, c2](y_10) at (18, 24) {};
\node[px, c2](x_10) at (19, 15) {};
\node[px, c3](x_11) at (20, 18) {};
\node[py, c3](y_11) at (21, 25) {};
\node[py, c0](y_12) at (22, 28) {};
\node[px, c0](x_12) at (23, 19) {};
\node[px, c1](x_13) at (24, 22) {};
\node[py, c1](y_13) at (25, 29) {};
\node[py, c2](y_14) at (26, 30) {};
\node[px, c2](x_14) at (27, 23) {};
\node[px, c3](x_15) at (28, 26) {};
\node[py, c3](y_15) at (29, 31) {};
\node[py, c0](y_16) at (30, 32) {};
\node[px, c0](x_16) at (31, 27) {};
\draw [c1] (y_1) -- (x_1);
\draw [black] (y_2) -- (y_1);
\draw [c2] (x_2) -- (y_2);
\draw [black] (x_3) -- (x_2);
\draw [c3] (y_3) -- (x_3);
\draw [black] (y_4) -- (y_3);
\draw [c0] (x_4) -- (y_4);
\draw [black] (x_5) -- (x_4);
\draw [c1] (y_5) -- (x_5);
\draw [black] (y_6) -- (y_5);
\draw [c2] (x_6) -- (y_6);
\draw [black] (x_7) -- (x_6);
\draw [c3] (y_7) -- (x_7);
\draw [black] (y_8) -- (y_7);
\draw [c0] (x_8) -- (y_8);
\draw [black] (x_9) -- (x_8);
\draw [c1] (y_9) -- (x_9);
\draw [black] (y_10) -- (y_9);
\draw [c2] (x_10) -- (y_10);
\draw [black] (x_11) -- (x_10);
\draw [c3] (y_11) -- (x_11);
\draw [black] (y_12) -- (y_11);
\draw [c0] (x_12) -- (y_12);
\draw [black] (x_13) -- (x_12);
\draw [c1] (y_13) -- (x_13);
\draw [black] (y_14) -- (y_13);
\draw [c2] (x_14) -- (y_14);
\draw [black] (x_15) -- (x_14);
\draw [c3] (y_15) -- (x_15);
\draw [black] (y_16) -- (y_15);
\draw [c0] (x_16) -- (y_16);
\draw [black, dash pattern={on 1.5pt off 1.5pt}] (x_2) -- (x_1);
\draw [black, dash pattern={on 1.5pt off 1.5pt}] (x_3) -- (x_2);
\draw [black, dash pattern={on 1.5pt off 1.5pt}] (x_4) -- (x_3);
\draw [black, dash pattern={on 1.5pt off 1.5pt}] (y_1) -- (x_4);
\draw [c1, dash pattern={on 1.5pt off 1.5pt}] (x_5) -- (y_1);
\draw [black, dash pattern={on 1.5pt off 1.5pt}] (x_6) -- (x_5);
\draw [c2, dash pattern={on 1.5pt off 1.5pt}] (y_2) -- (x_6);
\draw [black, dash pattern={on 1.5pt off 1.5pt}] (y_3) -- (y_2);
\draw [c3, dash pattern={on 1.5pt off 1.5pt}] (x_7) -- (y_3);
\draw [black, dash pattern={on 1.5pt off 1.5pt}] (x_8) -- (x_7);
\draw [c0, dash pattern={on 1.5pt off 1.5pt}] (y_4) -- (x_8);
\draw [black, dash pattern={on 1.5pt off 1.5pt}] (y_5) -- (y_4);
\draw [c1, dash pattern={on 1.5pt off 1.5pt}] (x_9) -- (y_5);
\draw [black, dash pattern={on 1.5pt off 1.5pt}] (x_10) -- (x_9);
\draw [c2, dash pattern={on 1.5pt off 1.5pt}] (y_6) -- (x_10);
\draw [black, dash pattern={on 1.5pt off 1.5pt}] (y_7) -- (y_6);
\draw [c3, dash pattern={on 1.5pt off 1.5pt}] (x_11) -- (y_7);
\draw [black, dash pattern={on 1.5pt off 1.5pt}] (x_12) -- (x_11);
\draw [c0, dash pattern={on 1.5pt off 1.5pt}] (y_8) -- (x_12);
\draw [black, dash pattern={on 1.5pt off 1.5pt}] (y_9) -- (y_8);
\draw [c1, dash pattern={on 1.5pt off 1.5pt}] (x_13) -- (y_9);
\draw [black, dash pattern={on 1.5pt off 1.5pt}] (x_14) -- (x_13);
\draw [c2, dash pattern={on 1.5pt off 1.5pt}] (y_10) -- (x_14);
\draw [black, dash pattern={on 1.5pt off 1.5pt}] (y_11) -- (y_10);
\draw [c3, dash pattern={on 1.5pt off 1.5pt}] (x_15) -- (y_11);
\draw [black, dash pattern={on 1.5pt off 1.5pt}] (x_16) -- (x_15);
\draw [c0, dash pattern={on 1.5pt off 1.5pt}] (y_12) -- (x_16);
\draw [black, dash pattern={on 1.5pt off 1.5pt}] (y_13) -- (y_12);
\draw [black, dash pattern={on 1.5pt off 1.5pt}] (y_14) -- (y_13);
\draw [black, dash pattern={on 1.5pt off 1.5pt}] (y_15) -- (y_14);
\draw [black, dash pattern={on 1.5pt off 1.5pt}] (y_16) -- (y_15);
	\end{tikzpicture}
	\hspace{5mm}
	\begin{tikzpicture}[
	scale={5/7},
		framed,
		baseline=(current bounding box.center),
		px/.style={fill, circle, inner sep=1.5pt},
		py/.style={fill, rectangle, inner sep=2pt},
		c0/.style=blue,
		c1/.style=red,
		c2/.style={black!50!green},
		c3/.style=brown]
		\node[px, c1](x_1) at (0, 0) {};
\node[py, c1](y_1) at (0, 1) {};
\node[py, c2](y_2) at (1, 1) {};
\node[px, c2](x_2) at (1, 0) {};
\node[px, c3](x_3) at (2, 0) {};
\node[py, c3](y_3) at (2, 1) {};
\node[py, c0](y_4) at (3, 1) {};
\node[px, c0](x_4) at (3, 0) {};
\node[px, c1](x_5) at (0, 2) {};
\node[py, c1](y_5) at (0, 3) {};
\node[py, c2](y_6) at (1, 3) {};
\node[px, c2](x_6) at (1, 2) {};
\node[px, c3](x_7) at (2, 2) {};
\node[py, c3](y_7) at (2, 3) {};
\node[py, c0](y_8) at (3, 3) {};
\node[px, c0](x_8) at (3, 2) {};
\node[px, c1](x_9) at (0, 4) {};
\node[py, c1](y_9) at (0, 5) {};
\node[py, c2](y_10) at (1, 5) {};
\node[px, c2](x_10) at (1, 4) {};
\node[px, c3](x_11) at (2, 4) {};
\node[py, c3](y_11) at (2, 5) {};
\node[py, c0](y_12) at (3, 5) {};
\node[px, c0](x_12) at (3, 4) {};
\node[px, c1](x_13) at (0, 6) {};
\node[py, c1](y_13) at (0, 7) {};
\node[py, c2](y_14) at (1, 7) {};
\node[px, c2](x_14) at (1, 6) {};
\node[px, c3](x_15) at (2, 6) {};
\node[py, c3](y_15) at (2, 7) {};
\node[py, c0](y_16) at (3, 7) {};
\node[px, c0](x_16) at (3, 6) {};
\draw [c1] (y_1) -- (x_1);
\draw [black] (y_2) -- (y_1);
\draw [c2] (x_2) -- (y_2);
\draw [black] (x_3) -- (x_2);
\draw [c3] (y_3) -- (x_3);
\draw [black] (y_4) -- (y_3);
\draw [c0] (x_4) -- (y_4);
\draw [black] (x_5) -- (x_4);
\draw [c1] (y_5) -- (x_5);
\draw [black] (y_6) -- (y_5);
\draw [c2] (x_6) -- (y_6);
\draw [black] (x_7) -- (x_6);
\draw [c3] (y_7) -- (x_7);
\draw [black] (y_8) -- (y_7);
\draw [c0] (x_8) -- (y_8);
\draw [black] (x_9) -- (x_8);
\draw [c1] (y_9) -- (x_9);
\draw [black] (y_10) -- (y_9);
\draw [c2] (x_10) -- (y_10);
\draw [black] (x_11) -- (x_10);
\draw [c3] (y_11) -- (x_11);
\draw [black] (y_12) -- (y_11);
\draw [c0] (x_12) -- (y_12);
\draw [black] (x_13) -- (x_12);
\draw [c1] (y_13) -- (x_13);
\draw [black] (y_14) -- (y_13);
\draw [c2] (x_14) -- (y_14);
\draw [black] (x_15) -- (x_14);
\draw [c3] (y_15) -- (x_15);
\draw [black] (y_16) -- (y_15);
\draw [c0] (x_16) -- (y_16);
\draw [black, dash pattern={on 1.5pt off 1.5pt}] (x_2) -- (x_1);
\draw [black, dash pattern={on 1.5pt off 1.5pt}] (x_3) -- (x_2);
\draw [black, dash pattern={on 1.5pt off 1.5pt}] (x_4) -- (x_3);
\draw [black, dash pattern={on 1.5pt off 1.5pt}] (y_1) -- (x_4);
\draw [c1, dash pattern={on 1.5pt off 1.5pt}] (x_5) -- (y_1);
\draw [black, dash pattern={on 1.5pt off 1.5pt}] (x_6) -- (x_5);
\draw [c2, dash pattern={on 1.5pt off 1.5pt}] (y_2) -- (x_6);
\draw [black, dash pattern={on 1.5pt off 1.5pt}] (y_3) -- (y_2);
\draw [c3, dash pattern={on 1.5pt off 1.5pt}] (x_7) -- (y_3);
\draw [black, dash pattern={on 1.5pt off 1.5pt}] (x_8) -- (x_7);
\draw [c0, dash pattern={on 1.5pt off 1.5pt}] (y_4) -- (x_8);
\draw [black, dash pattern={on 1.5pt off 1.5pt}] (y_5) -- (y_4);
\draw [c1, dash pattern={on 1.5pt off 1.5pt}] (x_9) -- (y_5);
\draw [black, dash pattern={on 1.5pt off 1.5pt}] (x_10) -- (x_9);
\draw [c2, dash pattern={on 1.5pt off 1.5pt}] (y_6) -- (x_10);
\draw [black, dash pattern={on 1.5pt off 1.5pt}] (y_7) -- (y_6);
\draw [c3, dash pattern={on 1.5pt off 1.5pt}] (x_11) -- (y_7);
\draw [black, dash pattern={on 1.5pt off 1.5pt}] (x_12) -- (x_11);
\draw [c0, dash pattern={on 1.5pt off 1.5pt}] (y_8) -- (x_12);
\draw [black, dash pattern={on 1.5pt off 1.5pt}] (y_9) -- (y_8);
\draw [c1, dash pattern={on 1.5pt off 1.5pt}] (x_13) -- (y_9);
\draw [black, dash pattern={on 1.5pt off 1.5pt}] (x_14) -- (x_13);
\draw [c2, dash pattern={on 1.5pt off 1.5pt}] (y_10) -- (x_14);
\draw [black, dash pattern={on 1.5pt off 1.5pt}] (y_11) -- (y_10);
\draw [c3, dash pattern={on 1.5pt off 1.5pt}] (x_15) -- (y_11);
\draw [black, dash pattern={on 1.5pt off 1.5pt}] (x_16) -- (x_15);
\draw [c0, dash pattern={on 1.5pt off 1.5pt}] (y_12) -- (x_16);
\draw [black, dash pattern={on 1.5pt off 1.5pt}] (y_13) -- (y_12);
\draw [black, dash pattern={on 1.5pt off 1.5pt}] (y_14) -- (y_13);
\draw [black, dash pattern={on 1.5pt off 1.5pt}] (y_15) -- (y_14);
\draw [black, dash pattern={on 1.5pt off 1.5pt}] (y_16) -- (y_15);
	\end{tikzpicture}
	\caption{A $2$-increasing permutation of length 32 whose incidence graph contains a $4 \times 8$ grid. Vertex shape indicates the track, colors are for emphasis of the grid structure.}
	\label{fig:bad_2track}
\end{figure}
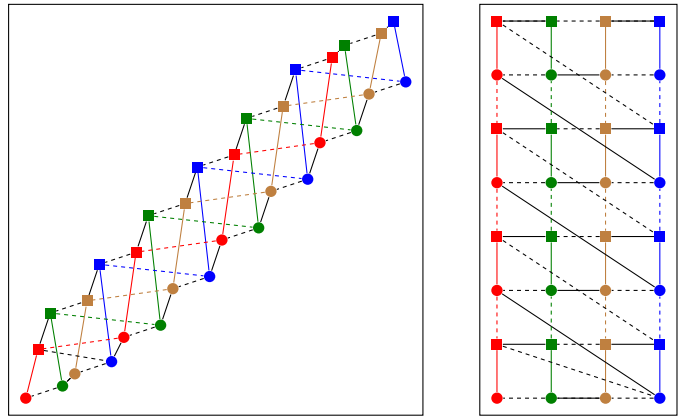

We now relabel the vertices to show the contained grid. For $i \in [k]$, $j \in [2k]$, let $z_{i,j} = x_{\lfloor j/2 \rfloor k + i}$ if $j$ is odd, and $z_{i,j} = y_{(j/2-1) k + i}$ if $j$ is even. It is easy to see that $z_{i,j}$ is adjacent to $z_{i+1,j}$ and $z_{i,j+1}$ for $i \in [k-1]$ and $j \in [2k-1]$. For an illustration of the obtained permutation and the contained grid graph, see Figure~\ref{fig:bad_2track}.

The case of Jordan patterns (Theorem~\ref{thmapp}(ii)) is immediate, as the incidence graph of Jordan permutations is by definition planar. We defer the details to Appendix~\ref{appd}.

\subparagraph*{$3$-monotone patterns.}

We now prove Theorem~\ref{thmembed}. We first need some definitions and observations. For a set $\Pi$ of length-$n$ permutations, define the graph $G_H(\Pi) = ([n], E)$, where $E$ is the union of the Hamiltonian paths corresponding to all $\pi \in \Pi$. For an arbitrary length-$n$ permutation $\pi$, the graph $G_\pi$ is isomorphic to $G_H(\{\mathrm{id}_n, \pi\})$, where $\mathrm{id}_n$ is the length-$n$ identity permutation.

Permutation $\pi'$ is a \emph{split} of a permutation $\pi$ if $\pi'$ arises from $\pi$ by moving a subsequence of $\pi$ to the front. For example, $(1,3,5,2,4)$ is a split of $\mathrm{id}_5$, obtained by moving $(1,3,5)$ to the front. We call a permutation \emph{split permutation} if it is a split of the identity permutation. Observe that for a length-$n$ split permutation $\sigma \neq \mathrm{id}_n$, there is a unique integer $p(\sigma) \in [n]$ such that both $\sigma(1), \sigma(2), \dots, \sigma(p(\sigma))$ and $\sigma(p(\sigma)+1), \sigma(p(\sigma)+2), \dots, \sigma(n)$ are increasing. Furthermore, $\sigma^{-1}$ is a merge of the two subsequences $1, 2, \dots, p(\sigma)$ and $p(\sigma) + 1, p(\sigma) + 2, \dots, n$.

If $\pi'$ is a split of $\pi$, then $\pi' = \pi \circ \sigma$ for some split permutation $\sigma$. Ahal and Rabinovich~\cite{Ahal} mention that every $n$-permutation can be obtained from $\mathrm{id}_n$ by at most $\lceil \log n \rceil$ splits. 
Let $\pi$ be an \emph{arbitrary} $n$-permutation and consider the sequence $\mathrm{id}_n = \pi_1, \dots, \pi_{m} = \pi$, where for each $i \in [m-1]$ we have $\pi_{i+1} = \pi_i \circ \sigma_i$ for some split permutation $\sigma_i \neq \mathrm{id}_n$, and $m \le \lceil \log n \rceil$. % for $i \in [m]$.
Let $\Pi= \{\pi_1, \dots, \pi_{m}\}$.

To prove Theorem~\ref{thmembed}, we show that the graph $G_H(\Pi)$ can be embedded (as a minor) in the incidence graph $G$ of some permutation of length at most $2 m n$.
%, where $m \le \lceil \log n \rceil$ and $\sigma_i \neq \mathrm{id}$ for $i \in [m]$. 
We further show that $G$ is a $3$-track graph (and thus, its underlying permutation can be assumed $3$-increasing). The lower bound on the treewidth of $G$ then follows by (i) choosing $\pi$ to be a permutation whose incidence graph has treewidth $\Omega(n)$, (ii) the fact that $G_\pi$ is a subgraph of $G_H(\Pi)$, and thus, a minor of $G$, and (iii) the observation that the treewidth of a graph is not less than the treewidth of its minor.

We first define the vertex sets corresponding to the three tracks of $G$. Let
\begin{align*}
& V_x = \{ x_{i,j} \mid i \in [m], j \in [n] \}, \\
& V_y = \{ y_{i,j} \mid i \in [m-1], j \in [p(\sigma_i)] \}, \text{ ~and} \\
& V_z = \{ z_{i,j} \mid i \in [m-1], j \in [n] \setminus [p(\sigma_i)] \}.
\end{align*}

Let $V = V_x \cup V_y \cup V_z$ be the vertex set of $G$, and observe that $|V| = mn + (m-1)n \le 2mn$. Figure \ref{fig:bad_3track} (left) shows the vertices in an arrangement useful for the rest of the construction.

To later show that $G$ is a $3$-track graph, we fix a total order $\prec$ on each track, namely, the lexicographic order of the vertex-indices, i.e.\ $x_{i,j} \prec x_{i',j'}$ if and only if $i < i'$ or $(i = i') \wedge (j < j')$, and analogously for $V_y$ and $V_z$. Before proceeding, we define the following functions:
\begin{align*}
& s_x \colon V_x \setminus \{x_{m,n}\} \rightarrow V_x \setminus \{x_{1,1}\}, \\
& s_x(x_{i,j}) = \begin{cases}
x_{i,j+1}, & \text{ if } j < n, \\
x_{i+1,1}, & \text{ if } j = n.
\end{cases} \\
& s_c \colon V \setminus \{x_{m, j} \mid j \in [n] \} \rightarrow V \setminus \{x_{1,j} \mid j \in [n]\}, \\
& s_c(x_{i, j}) = \begin{cases}
y_{i, \sigma^{-1}_i(j)}, & \text{ if } \sigma^{-1}_i(j) \le p(\sigma_i), \\
z_{i, \sigma^{-1}_i(j)}, & \text{ if } \sigma^{-1}_i(j) > p(\sigma_i). \\
\end{cases} \\
& s_c(y_{i,j}) = x_{i+1, j}, \\
& s_c(z_{i,j}) = x_{i+1, j}.
\end{align*}

Note that $s_x$ is just the successor with respect to the total order $\prec$ on $V_x$, and that $s_c$ is a bijection. \Cref{fig:bad_3track} (middle) illustrates the two functions.

Now we define the two Hamiltonian paths whose union is $G$. The first path $P_1$ goes as follows: start at $x_{1,1}$, then, from every $x_{i,j}$ with $i < m$, go to $s_c( x_{i,j} )$, and then to $s_x( x_{i,j} )$. For $x_{m,j}$ with $j < n$, go directly to $s_x(x_{m,j})$. Path $P_1$ contains all vertices of $V_x$ in the correct order. The same holds for $V_y$ and $V_z$, by the definition of $s_c$.

The second path $P_2$ also starts at $x_{1,1}$, but first goes along $V_x$ until it reaches $x_{2,1}$, i.e.\ the first part of $P_2$ is $x_{1,1}, x_{1,2}, \dots, x_{1,n}, x_{2,1}$. Then, from every $x_{i,j}$ with $i \ge 2$, it first moves to $s^{-1}_c(x_{i,j})$ and then to $s_x(x_{i,j})$. Again, $P_2$ contains all vertices of $V_x$ in the correct order. As $s^{-1}_c(x_{i,j})$ is either $y_{i-1,j}$ or $z_{i-1,j}$, this is also true for $V_y$ and $V_z$.

To obtain $G_H(\Pi) = ([n], E)$, color the vertices of the graph with $n$ colors, where color $k$ induces a path $C_k$ of length $m$ in $G$. We then prove that for each $\{k_1, k_2\} \in E$, the graph $G$ contains adjacent vertices of the colors $k_1$ and $k_2$. Then, by contracting $C_k$ for $k \in [n]$, we obtain a supergraph of $G_H(\Pi)$. See \cref{fig:bad_3track} (right) for an illustration.

For $k \in [n]$, define the path $C_k = ( x_{1,k}, s_c(x_{1,k}), s^2_c(x_{1,k}), \dots, s^{2m-2}_c(x_{1,k}) )$. As $s_c$ is a bijection, these paths are disjoint. Note that for each $x_{i,j} \in V_x \setminus \{x_{m,n}\}$,
\begin{align*}
& s^2_c(x_{i,j}) = x_{i+1, \sigma^{-1}_i(j)}.
\end{align*}
We claim that the color of $x_{i,j}$ is $\pi_i(j)$. This is because:
\begin{align*}
s_c^{2i-2}( x_{1,\pi_i(j)}) & = s_c^{2i-2}( x_{1,\sigma_1 \sigma_2 \dots \sigma_{i-1}(j)}) \\
& = s_c^{2i-4}( x_{2, \sigma^{-1}_1 \sigma_1 \sigma_2 \dots \sigma_i(j)}) = s_c^{2i-4}( x_{2, \sigma_2 \sigma_3 \dots \sigma_{i-1}(j)} ) \\
& = ... = s_c^{2i - 2\ell}( x_{\ell, \sigma_\ell \sigma_{\ell+1} \dots \sigma_{i-1}(j)} ) \\
& = ... = x_{i,j}. 
\end{align*}

Now let $k_1$ and $k_2$ be adjacent in $G_H(\Pi)$. Then, there exist $i, j$ such that $\pi_i(j) = k_1$ and $\pi_i(j+1) = k_2$ and, as discussed above, $x_{i,j} \in C_{k_1}$ and $x_{i,j+1} \in C_{k_2}$. By definition $x_{i,j} \in C_{k_1}$ implies $s_c^{-1}(x_{i,j}) \in C_{k_1}$. Finally, $P_2$ has an edge from $s_c^{-1}(x_{i,j})$ to $s_x(x_{i,j}) = x_{i,j+1}$. This concludes the proof.

\medskip

The construction can be extended to embed the union of $k$ arbitrary Hamiltonian paths on $n$ vertices as a minor of a $3$-track graph with ${O}(k n \log n)$ vertices. As every order-$n$ graph of maximum degree $d$ is edge-colorable with $d+1$ colors (by Vizing's theorem), such graphs are in the union of at most $d+1$ Hamiltonian paths, can thus be embedded in $3$-track graphs of order ${O}(d n \log{n})$. 

\section{Hardness result} \label{sec:hardness}

In this section we prove Theorem~\ref{thmhard}. The hardness proof proceeds in two steps. First, we reduce the \emph{partitioned subgraph isomorphism} (PSI) problem to the \emph{partitioned permutation pattern matching} (PPPM) problem. 
Then, we reduce from the more difficult, counting variant of PPPM to the regular counting PPM (the subject of Theorem~\ref{thmhard}), using a (by now standard) technique based on inclusion-exclusion.

\subparagraph*{PSI to PPPM.}
The input to the PSI problem (introduced in~\cite{MarxTW}) consists of a graph $G$, a graph $H$, and a coloring $\phi$ of $V(G)$ with colors $V(H)$. The task is to decide whether there is a mapping $g : V(H) \rightarrow V(G)$ such that $\{u,v\} \in E(H)$ if and only if $\{g(u),g(v)\} \in E(G)$, and $\phi(g(u)) = u$ for all $u \in V(H)$. In words, we look for a subgraph of $G$ that is isomorphic to $H$, with the restriction that each vertex of $H$ can only correspond to a vertex of $G$ from a prescribed set, moreover, these sets are disjoint.

Let $n$ denote the number of vertices of $G$, and let $k$ denote the number of \emph{edges} of $H$. It is known~\cite[Corr.\ 6.3]{MarxTW}, that PSI cannot be solved in time $f(k) \cdot n^{o(k/\log{k})}$, unless ETH fails, moreover, this holds even if $|E(H)| = |V(H)|$ (see e.g.\ \cite{HSVC}).

The input to the PPPM problem (introduced in~\cite{GM}) consists of permutations $\tau$ and $\pi$ of lengths $n$ and $k$ respectively, and a coloring $\phi: [n] \rightarrow [k]$ of the entries of $\tau$. The task is to decide whether there is an embedding $g : [k] \rightarrow [n]$ of $\pi$ into $\tau$ in the sense of the standard PPM problem, with the additional restriction that $\phi(g(i)) = i$, for all $i \in [k]$.

Guillemot and Marx show~\cite[Thm.\ 6.1]{GM}, through a reduction from \emph{partitioned clique}, that PPPM is $W[1]$-hard. Due to the density of a clique, the same reduction would, at best, yield a lower bound with exponent $\sqrt{k}$. We strengthen (and somewhat simplify) this reduction, to show that PPPM is at least hard as PSI, obtaining the following.

\begin{lemma}[Proof in Appendix~\ref{appf}]\label{lemhard}
PPPM cannot be solved in time $f(k) \cdot n^{o(k/\log{k})}$, unless ETH fails. 
\end{lemma}

\subparagraph*{\#PPPM to \#PPM.}

The counting variant of PPPM (denoted \#PPPM) is clearly at least as hard as PPPM. We now show that the counting variant of PPM (denoted \#PPM) is at least as hard as \#PPPM, thereby proving Theorem~\ref{thmhard}.

We use oracle-calls to \#PPM for all subsets $X \subseteq [k]$, to count the number of embeddings of $\pi$ into $\tau$ using entries of $\tau$ with colors from the set $X$, but ignoring colors for the purpose of the embedding. (We can achieve this by deleting the entries of $\tau$ with color in $[k] \setminus X$ before each oracle-call.) 
Then, using the inclusion-exclusion formula, we obtain the number $C$ of embeddings that use \emph{all} colors in $[k]$ as follows: 

$$
C = \sum_{X \subseteq [k]}{(-1)^{k - |X|} ~C_X},
$$

where $C_X$ denotes the number of embeddings that use colors from the set $X$ (obtained by oracle calls). Since $\pi$ is of length $k$, the quantity $C$ counts exactly the number of embeddings that use each color once. 

It remains to show that embeddings that use every color in $[k]$ are such that $\pi_i$ is matched to an entry of $\tau$ of color $i$, for all $i \in [k]$, i.e.\ the colors are not permuted. This is indeed the case for the hard instance constructed in the proof of Lemma~\ref{lemhard}. Towards this claim (referring to Appendix~\ref{appf}) observe that all points that are unique in their respective pattern-cell $(i,j)$ can only be matched to a point in the corresponding text-cell $(i,j)$, which is of the correct color. In each diagonal cell $(i,i)$, for $i > 0$, there is a matched point, and the pattern has two bracketing points in decreasing order in pattern-cell $(i,0)$, and two bracketing points in increasing order in pattern-cell $(0,i)$. By construction, the only two points in the correct order are the nearest bracketing points in text-cell $(i,0)$, resp.\ $(0,i)$, which are indeed of the correct color. 

The number of oracle calls and additional overhead amounts to a factor $2^k$ in the running time, absorbed in the quantity $f(k) \cdot n^{o(k/\log{k})}$. This concludes the proof.

\medskip

\subparagraph*{Acknowledgements.}

An earlier version of the paper contained a mistake in the analysis of the algorithm for Theorem~\ref{thm2}. We thank G\"unter Rote for pointing out the error.

This work was prompted by the Dagstuhl Seminar 18451 ``Genomics, Pattern Avoidance, and Statistical Mechanics''. The second author thanks the organizers for the invitation and the participants for interesting discussions.

\newpage
\appendix

\section{Appendix}
\subsection{Proof of Lemma~\ref{eqlem}\label{appa}}
\begin{proof}
Suppose $\tau$ contains $\pi$, and let $(\tau(i_1),\dots,\tau(i_k))$ be the subsequence witnessing this fact. Let $p_j$ denote the point $(j,\pi(j))$, and set $f(p_j) = (i_j, \tau(i_j))$ for all $j \in [k]$. Observe that $f(N^L(p_j)).x = i_{j-1}$, and $f(N^R(p_j)).x = i_{j+1}$, the first condition thus holds since $i_{j-1} < i_j < i_{j+1}$. 

Let $\pi(j') = N^D(p_j).y$, and $\pi(j'') = N^U(p_j).y$. By definition, $\pi(j') < \pi(j) < \pi(j'')$. The second condition now becomes $\tau(i_{j'}) < \tau(i_j) < \tau(i_{j''})$, which holds since $\tau$ contains $\pi$.

In the other direction, let $f:S_\pi \rightarrow S_\tau$ be a valid embedding.
Define $i_j = f(p_j).x$, for all $j\in[k]$. Since $f(N^L(p_j)).x < f(p_j).x < f(N^R(p_j)).x$ for all $j$, we have $i_1 \leq \cdots \leq i_k$. Let $j',j'' \in [k]$, such that $j' < j''$. 

Then $\pi(i_{j'}) < \pi(i_{j''})$ is equivalent with $\tau(i_{j'}) = f(p_j).y < f(N^U(\dots(N^U(p_j))\dots)).y = \tau(i_{j''})$ where the $N^U(\cdot)$ operator, and the second property of a valid embedding are applied $j'' - j'$ times. %The case $j'>j''$ is symmetric.
\end{proof}

\subsection{The even-odd method \label{appb}}
In this section we describe the algorithm of Theorem~\ref{thm2} in a self-contained manner, \emph{not using} the formalism of CSPs.

Let $(Q^E, Q^O)$ be the partition of $S_\pi$ into points with even and odd indices. Formally, $Q^{E} = \{(2k, \pi(2k)) \mid 1 \leq k \leq \lfloor k/2 \rfloor \}$, and $Q^{O} = \{(2k-1,\pi(2k-1)) \mid 1 \leq k \leq \lceil k/2 \rceil \}$. 

Suppose $\tau$ contains $\pi$. Then, by Lemma~\ref{eqlem}, there exists a valid embedding $f:S_\pi \rightarrow S_\tau$. 
We start by \emph{guessing} a partial embedding $g_0: Q^{E} \rightarrow S_\tau$. (For example, $g_0 = f|_{Q^E}$ is such a partial embedding.)
We then extend $g_{0}$ step-by-step, adding points to its domain, until it becomes a valid embedding $S_\pi \rightarrow S_\tau$. 

Let $p_1, \dots, p_{\lceil k/2 \rceil}$ be the elements of $Q^{O}$ in increasing order of \emph{value}, i.e.\ $1 \leq p_1.y < \cdots < p_{\lceil k/2 \rceil}.y \leq n$, and let $P_0 = \emptyset$ and $P_i = P_{i-1} \cup \{p_i\}$, for $1 \leq i \leq \lceil k/2 \rceil$. For all $i$, we maintain the invariant that $g_i$ is a restriction of some valid embedding to $Q^{E} \cup P_i$. By our choice of $g_0$, this is true initially for $i=0$.

In the $i$-th step (for $i=0,\dots,\lceil k/2 \rceil-1$), we extend $g_i$ to $g_{i+1}$ by mapping the next point $p_{i+1}$ onto a suitable point in $S_\tau$. For $g_{i+1}$ to be a restriction of a valid embedding, it must satisfy conditions (1) and (2) on the relative position of neighbors. Observe that all, except possibly one, of the neighbors of $p_{i+1}$ are already embedded by $g_i$. This is because $N^L(p_{i+1})$ and $N^R(p_{i+1})$ have even index, are thus in $Q^E$, unless they are virtual points and thus implicitly embedded. The point $N^D(p_{i+1})$ is either an even-index point, and thus in $Q^E$, or the virtual point $(0,0)$ and thus implicitly embedded, or an odd-index point, in which case, by our ordering, it must be $p_i$, and thus, contained in $P_i$. The only neighbor of $p_{i+1}$ possibly not embedded is $N^U(p_{i+1})$. 

If we map $p_{i+1}$ to a point $q \in S_\tau$, we have to observe the constraints $g_{i}(N^{L}(p_{i+1})).x < q.x < g_{i}(N^{R}(p_{i+1})).x$, and $g_{i}(N^{D}(p_{i+1})).y < q.y$. If $N^U(p_{i+1})$ is also in the domain of $g_i$, then we have the additional constraint $q.y < g_{i}(N^{U}(p_{i+1})).y$.

These constraints determine an (open) axis-parallel box, possibly extending upwards infinitely (in case only three of the four neighbors of $p_{i+1}$ are embedded so far). Assuming $g_i$ is a restriction of a valid embedding $f$, the point $f(p_{i+1})$ must satisfy all constraints, it is thus contained in this box. We extend $g_i$ to obtain $g_{i+1}$ by mapping $p_{i+1}$ to a point $q \in S_\tau$ in the constraint-box, and if there are multiple such points, we pick the one that is \emph{lowest}, i.e.\ the one with smallest value $q.y$. 

The crucial observation is that if $g_i$ is a partial embedding, then $g_{i+1}$ is also a partial embedding, and the correctness of the procedure follows by induction.

Indeed, some valid embedding $f' : S_\pi \rightarrow S_\tau$ must be the extension of $g_{i+1}$. If $q = f(p_{i+1})$, then $f'$ is $f$ itself. Otherwise, let $f'$ be identical with $f$, except for mapping $p_{i+1} \rightarrow q$ (instead of mapping $p_{i+1} \rightarrow f(p_{i+1})$). The only conditions of a valid embedding that may become violated are those involving $p_{i+1}$. The conditions $f'(N^L(p_{i+1})).x < f'(p_{i+1}).x < f'(N^R(p_{i+1})).x$ and $f'(N^D(p_{i+1})).y < f'(p_{i+1}).y$ hold by our choice of $q$.

\begin{figure*}
  \centering
  \includegraphics[width=12cm]{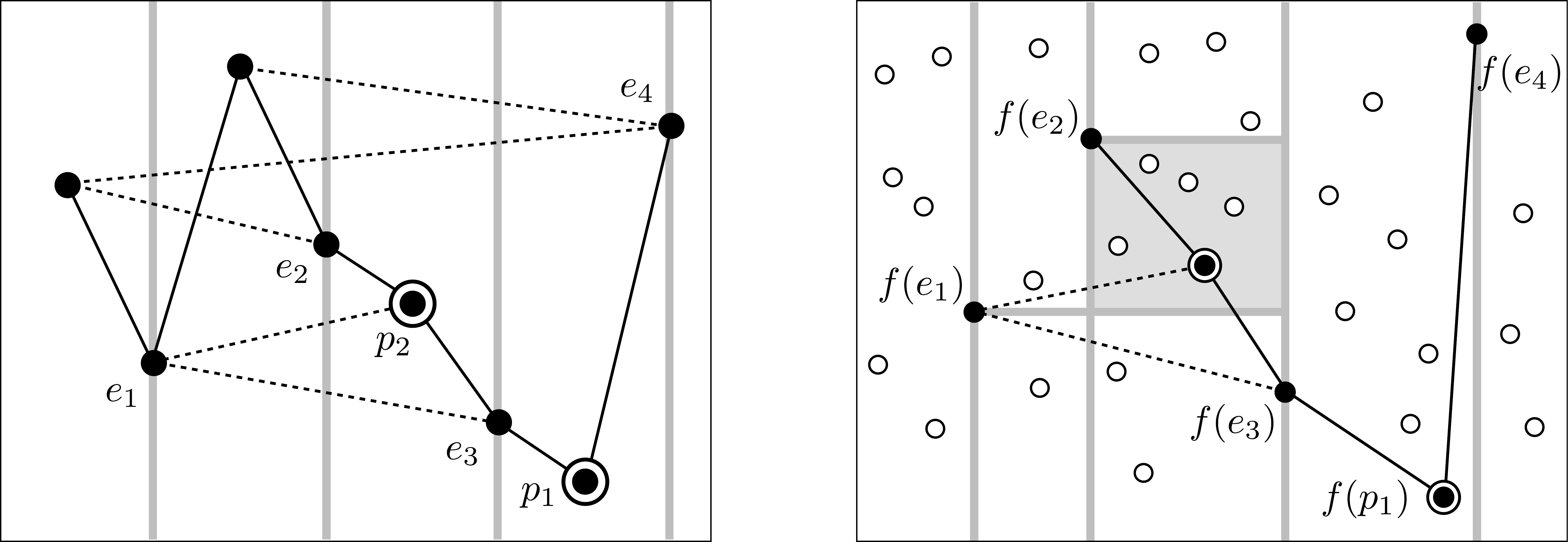}
  \caption{(\emph{left}) Pattern $\pi = (6,3,8,5,4,2,1,7)$ and its incidence graph $G_\pi$. Solid lines indicate neighbors by index, dashed lines indicate neighbors by value (lines may overlap). (\emph{right}) Text permutation $\tau$, points shown as circles. Partial embedding of $\pi$ shown with filled circles. Vertical bars mark even-index points $e_1$, $e_2$, $e_3$, $e_4$. Double circles mark the first two odd-index points $p_1$, $p_2$. Shaded box indicates constraints for embedding $p_{2}$, determined by $N^U(p_{2}) = N^L(p_2) = e_2$, $N^D(p_{2}) = e_1$, and $N^R(p_{2}) = e_3$. Observe that $p_2$ is mapped to lowest point (by value) that satisfies constraints. Revealed edges of $G_\pi$ are shown.  \label{algo}}
\end{figure*}

The condition $f'(p_{i+1}).y < f'(N^U(p_{i+1})).y$ holds a fortiori since we picked the \emph{lowest point} in a box that also contained $f(p_{i+1})$, in other words, $f'(p_{i+1}).y = q.y \leq f(p_{i+1}).y < f(N^U(p_{i+1})).y = f'(N^U(p_{i+1})).y$. Thus, $g_{i+1}$ is a partial embedding, which concludes the argument. See Figure~\ref{algo} for illustration.

Assuming that our initial guess $g_{0}$ was correct, we succeed in constructing a valid embedding that certifies the fact that $\tau$ contains $\pi$. We remark that guessing $g_0$ should be understood as trying all possible embeddings of $Q^E$. If our choice of $g_0$ is incorrect, i.e.\ not a partial embedding, then we reach a situation where we cannot extend $g_i$, and we abandon the choice of $g_0$. If extending $g_0$ to a valid embedding fails for all initial choices, we conclude that $\tau$ does not contain $\pi$. The resulting algorithm is described in Figure~\ref{algoS}. 

\begin{figure*}
%\begin{center}
\begin{algorithm}[H]
\DontPrintSemicolon
\NoCaptionOfAlgo
\caption{Polynomial-space algorithm for PPM:}

\For{\textbf{\emph{all~}} $g_0: Q^E \rightarrow S_t$}{
\textbf{if} $g_0$ not valid, \textbf{then next} $g_0$ \;
\For{$i\gets 0$ \KwTo $ \lceil k/2 \rceil - 1$}{
    \textbf{let} $q \in S_\tau$ with minimum $q.y$ such that: \;
    ~~~~$g_i(N^L(p_{i+1})).x < q.x < g_i(N^R(p_{i+1})).x$ \; 
    ~~~~$g_i(N^D(p_{i+1})).y < q.y$ \; 
    ~~~~$g_i(N^U(p_{i+1})).y> q.y$ ~~~(in case $N^U(p_{i+1}) \in Q^E$) \;
\textbf{if} no such $q$, \textbf{then next} $g_0$ \;
\textbf{extend} $g_i$ to $g_{i+1}$ by mapping $p_{i+1} \rightarrow q$ \;
}
\textbf{return} $g_{\lceil k/2 \rceil}$
}
\textbf{return} ``$\tau$ avoids $\pi$''
\end{algorithm}
\caption{\label{algoS}Finding a valid embedding of $S_\pi$ into $S_t$, or reporting that $t$ avoids $\pi$, with precomputed $Q^E$ (even-index points of $S_\pi$) and $(p_1, \dots, p_{\lceil k/2 \rceil})$ (odd-index points of $S_\pi$ sorted by value). \;}
%
%\end{center}
\end{figure*}

The space requirement is linear in the input size; apart from minor bookkeeping, only a single embedding must be stored at all times.
To analyse the running time, observe first, that $g_0$ must map points in $Q^E$ to points in $S_\tau$, preserving their left-to-right order (by index), and their bottom-to-top order (by value). The first condition can be enforced directly, by considering only \emph{subsequences} of $\tau$. This amounts to ${n \choose \lfloor k/2 \rfloor}$ choices for $g_0$ in the outer loop. The second condition can be verified in a linear time traversal of $G_\pi$ (this is the second line of the algorithm in Figure~\ref{algoS}). 

All remaining steps can be  performed using straightforward data structuring: we need to traverse to neighbors in the incidence graph, to go from $x$ to $g_i(x)$ and back, and to answer rectangle-minimum queries; all can be achieved in constant time, with a polynomial time preprocessing. We can in fact do away with rectangle queries, since candidate points of $\tau$ are considered in increasing order of value---the inner loop thus consists of a single sweep through both $\pi$ and $\tau$, which can be implemented in $O(n)$ time. By a standard bound on the binomial coefficient, the claimed running time of $O(n^{k/2 + 1})$ follows.

The efficient enumeration of initial embeddings with the required property can be achieved with standard techniques, see e.g.~\cite{comb_book}.  
Finally, we remark that instead of trying all embeddings $g_0$, in practice it may be preferable to build such an embedding incrementally, using backtracking. This allows the process to ``fail early'' if a certain embedding can not be extended to any partial embedding of $Q^E$.  
The order in which points of $Q^E$ are considered in the backtracking process can affect the performance significantly, see~\cite{knuth1975estim,knuth2000dancing} for consideration of similar issues. Alternatively, a modification of the algorithm of Theorem~\ref{thm1} may also be used to enumerate all valid initial $g_0$.

\subsection{Jordan patterns\label{appd}}

The proof of Theorem~\ref{thmapp}(ii) is immediate: the incidence graph of Jordan permutations is by definition planar. To see this, recall that a Jordan permutation is defined by the intersection pattern of two curves. We view the curves as the planar embedding of $G_\pi$. The portions of the curves between intersection points correspond to edges (we trim away the loose ends of both curves), and the curves connect the points in the order of their index, resp.\ value. This turns out to be an exact characterization: $G_\pi$ is planar if and only if $\pi$ is a Jordan permutation.

We remark that for the ``only if'' direction to hold, \emph{touching points} between the two curves must also be allowed. Consider any noncrossing embedding of $G_\pi$, and construct the two curves as the Hamiltonian paths of $G_\pi$ that connect the vertices by increasing index, resp.\ value. Whenever the two curves overlap over an edge of $G_\pi$, bend the corresponding part of one of the curves, such as to create two intersection points at the two endpoints of the edge (one of the two intersection points may need to be a touching point).

\medskip

The tightness of the result follows by the example of §\,\ref{sec:spec}. As $2$-track graphs are planar, the given $2$-increasing permutation (whose incidence-graph contains a large grid) is also a Jordan-permutation. 

\subsection{Treewidth-bound for $\sigma$-avoiding patterns \label{appq}}

In §\,\ref{sec:spec} we constructed length-$k$ permutations $\pi$ that avoid $(1,2,3)$ or $(3,2,1)$ with the property that $\tw(\pi) = \Omega(\sqrt{k})$. The same construction works for length-$k$ permutations that avoid an arbitrary $\sigma$, for $|\sigma| \geq 5$: by the Erd\H{o}s-Szekeres theorem all permutations of length $5$ or more contain $(1,2,3)$ or $(3,2,1)$, therefore, avoiding $(1,2,3)$ or $(3,2,1)$ implies avoiding $\sigma$. 

The only length-$4$ permutations $\sigma$ that avoid both $(1,2,3)$ and $(3,2,1)$ are $(2,1,4,3)$, $(3,1,4,2)$, and their reverses. 
In these cases we can construct a large-treewidth $\sigma$-avoiding permutation using a structural observation of Jel\'{i}nek and Kyn\u{c}l~\cite{hardness321}. They show that permutations that avoid \emph{both} $(2,1,4,3)$ and $(3,1,4,2)$ have a certain spiraling block-decomposition (see~\cite[Fig.\ 13]{hardness321}). Given this structure, the embedding of a grid is a straightforward adaptation of the technique in §\,\ref{sec:spec} (we omit the details).

It follows that there exist $\sigma$-avoiding length-$k$ patterns $\pi$ with treewidth $\Omega(\sqrt{k})$ for all patterns $\sigma$, except when $\sigma \in \left\{(1), (1,2), (2,1), (1,3,2), (2,3,1), (2,1,3), (3,1,2)\right\}$, i.e.\ exactly the cases when PPM is polynomial-time solvable~\cite{hardness321}.

\subsection{Illustration for the proof of Theorem~\ref{thmembed} ($3$-increasing permutation) \label{appe}}
\begin{figure}[h]
	\centering
	
	\begin{tikzpicture}[
		baseline=0,
		framed,
		vertex/.style={circle, fill, inner sep=1pt}
	]
	\input{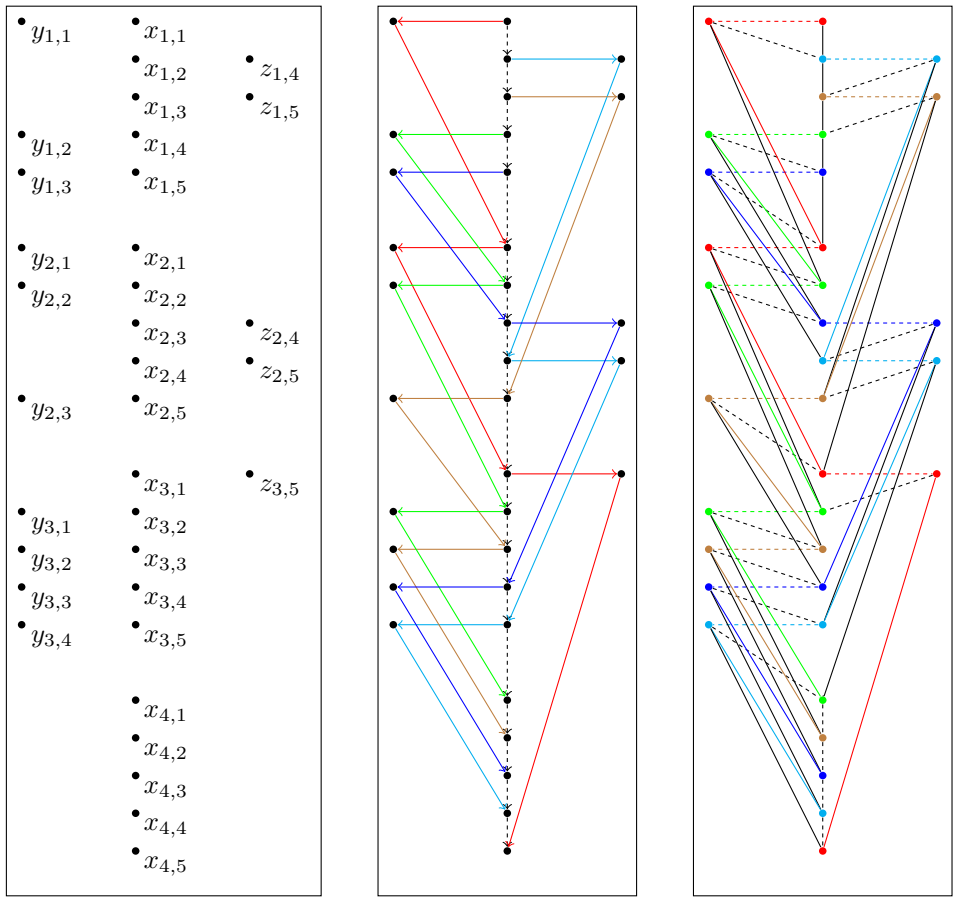}
	\coordinate (firstn) at (current bounding box.north);
	\coordinate (firsts) at (current bounding box.south);
	\end{tikzpicture}
	\hspace{5mm}
	\begin{tikzpicture}[
		baseline=0,
		framed,
		vertex/.style={circle, fill, inner sep=1pt},
		path_1/.style={->, dash pattern={on 1.5pt off 1.5pt}},
		path_2/.style={->}
	]
	\path (firstn) rectangle (firsts); % Enlarge height of bounding box with invisible path
	\node[vertex](x_1_1) at (0.0, 0.0) {};
\node[vertex](y_1_1) at (-1.5, 0.0) {};
\node[vertex](x_1_2) at (0.0, -0.5) {};
\node[vertex](z_1_4) at (1.5, -0.5) {};
\node[vertex](x_1_3) at (0.0, -1.0) {};
\node[vertex](z_1_5) at (1.5, -1.0) {};
\node[vertex](x_1_4) at (0.0, -1.5) {};
\node[vertex](y_1_2) at (-1.5, -1.5) {};
\node[vertex](x_1_5) at (0.0, -2.0) {};
\node[vertex](y_1_3) at (-1.5, -2.0) {};
\node[vertex](x_2_1) at (0.0, -3.0) {};
\node[vertex](y_2_1) at (-1.5, -3.0) {};
\node[vertex](x_2_2) at (0.0, -3.5) {};
\node[vertex](y_2_2) at (-1.5, -3.5) {};
\node[vertex](x_2_3) at (0.0, -4.0) {};
\node[vertex](z_2_4) at (1.5, -4.0) {};
\node[vertex](x_2_4) at (0.0, -4.5) {};
\node[vertex](z_2_5) at (1.5, -4.5) {};
\node[vertex](x_2_5) at (0.0, -5.0) {};
\node[vertex](y_2_3) at (-1.5, -5.0) {};
\node[vertex](x_3_1) at (0.0, -6.0) {};
\node[vertex](z_3_5) at (1.5, -6.0) {};
\node[vertex](x_3_2) at (0.0, -6.5) {};
\node[vertex](y_3_1) at (-1.5, -6.5) {};
\node[vertex](x_3_3) at (0.0, -7.0) {};
\node[vertex](y_3_2) at (-1.5, -7.0) {};
\node[vertex](x_3_4) at (0.0, -7.5) {};
\node[vertex](y_3_3) at (-1.5, -7.5) {};
\node[vertex](x_3_5) at (0.0, -8.0) {};
\node[vertex](y_3_4) at (-1.5, -8.0) {};
\node[vertex](x_4_1) at (0.0, -9.0) {};
\node[vertex](x_4_2) at (0.0, -9.5) {};
\node[vertex](x_4_3) at (0.0, -10.0) {};
\node[vertex](x_4_4) at (0.0, -10.5) {};
\node[vertex](x_4_5) at (0.0, -11.0) {};
\draw [path_1] (x_1_1) -- (x_1_2);
\draw [path_1] (x_1_2) -- (x_1_3);
\draw [path_1] (x_1_3) -- (x_1_4);
\draw [path_1] (x_1_4) -- (x_1_5);
\draw [path_1] (x_1_5) -- (x_2_1);
\draw [path_1] (x_2_1) -- (x_2_2);
\draw [path_1] (x_2_2) -- (x_2_3);
\draw [path_1] (x_2_3) -- (x_2_4);
\draw [path_1] (x_2_4) -- (x_2_5);
\draw [path_1] (x_2_5) -- (x_3_1);
\draw [path_1] (x_3_1) -- (x_3_2);
\draw [path_1] (x_3_2) -- (x_3_3);
\draw [path_1] (x_3_3) -- (x_3_4);
\draw [path_1] (x_3_4) -- (x_3_5);
\draw [path_1] (x_3_5) -- (x_4_1);
\draw [path_1] (x_4_1) -- (x_4_2);
\draw [path_1] (x_4_2) -- (x_4_3);
\draw [path_1] (x_4_3) -- (x_4_4);
\draw [path_1] (x_4_4) -- (x_4_5);
\draw [path_2, red] (x_1_1) -- (y_1_1);
\draw [path_2, red] (y_1_1) -- (x_2_1);
\draw [path_2, red] (x_2_1) -- (y_2_1);
\draw [path_2, red] (y_2_1) -- (x_3_1);
\draw [path_2, red] (x_3_1) -- (z_3_5);
\draw [path_2, red] (z_3_5) -- (x_4_5);
\draw [path_2, cyan] (x_1_2) -- (z_1_4);
\draw [path_2, cyan] (z_1_4) -- (x_2_4);
\draw [path_2, cyan] (x_2_4) -- (z_2_5);
\draw [path_2, cyan] (z_2_5) -- (x_3_5);
\draw [path_2, cyan] (x_3_5) -- (y_3_4);
\draw [path_2, cyan] (y_3_4) -- (x_4_4);
\draw [path_2, brown] (x_1_3) -- (z_1_5);
\draw [path_2, brown] (z_1_5) -- (x_2_5);
\draw [path_2, brown] (x_2_5) -- (y_2_3);
\draw [path_2, brown] (y_2_3) -- (x_3_3);
\draw [path_2, brown] (x_3_3) -- (y_3_2);
\draw [path_2, brown] (y_3_2) -- (x_4_2);
\draw [path_2, green] (x_1_4) -- (y_1_2);
\draw [path_2, green] (y_1_2) -- (x_2_2);
\draw [path_2, green] (x_2_2) -- (y_2_2);
\draw [path_2, green] (y_2_2) -- (x_3_2);
\draw [path_2, green] (x_3_2) -- (y_3_1);
\draw [path_2, green] (y_3_1) -- (x_4_1);
\draw [path_2, blue] (x_1_5) -- (y_1_3);
\draw [path_2, blue] (y_1_3) -- (x_2_3);
\draw [path_2, blue] (x_2_3) -- (z_2_4);
\draw [path_2, blue] (z_2_4) -- (x_3_4);
\draw [path_2, blue] (x_3_4) -- (y_3_3);
\draw [path_2, blue] (y_3_3) -- (x_4_3);
	\end{tikzpicture}
	\hspace{5mm}
	\begin{tikzpicture}[
		baseline=0,
		framed,
		vertex/.style={circle, fill, inner sep=1pt},
		p1/.style={dash pattern={on 1.5pt off 1.5pt}},
		p2/.style={}
	]
	\path (firstn) rectangle (firsts); % Enlarge height of bounding box with invisible path
	\node[vertex, red](x_1_1) at (0.0, 0.0) {};
\node[vertex, red](y_1_1) at (-1.5, 0.0) {};
\node[vertex, cyan](x_1_2) at (0.0, -0.5) {};
\node[vertex, cyan](z_1_4) at (1.5, -0.5) {};
\node[vertex, brown](x_1_3) at (0.0, -1.0) {};
\node[vertex, brown](z_1_5) at (1.5, -1.0) {};
\node[vertex, green](x_1_4) at (0.0, -1.5) {};
\node[vertex, green](y_1_2) at (-1.5, -1.5) {};
\node[vertex, blue](x_1_5) at (0.0, -2.0) {};
\node[vertex, blue](y_1_3) at (-1.5, -2.0) {};
\node[vertex, red](x_2_1) at (0.0, -3.0) {};
\node[vertex, red](y_2_1) at (-1.5, -3.0) {};
\node[vertex, green](x_2_2) at (0.0, -3.5) {};
\node[vertex, green](y_2_2) at (-1.5, -3.5) {};
\node[vertex, blue](x_2_3) at (0.0, -4.0) {};
\node[vertex, blue](z_2_4) at (1.5, -4.0) {};
\node[vertex, cyan](x_2_4) at (0.0, -4.5) {};
\node[vertex, cyan](z_2_5) at (1.5, -4.5) {};
\node[vertex, brown](x_2_5) at (0.0, -5.0) {};
\node[vertex, brown](y_2_3) at (-1.5, -5.0) {};
\node[vertex, red](x_3_1) at (0.0, -6.0) {};
\node[vertex, red](z_3_5) at (1.5, -6.0) {};
\node[vertex, green](x_3_2) at (0.0, -6.5) {};
\node[vertex, green](y_3_1) at (-1.5, -6.5) {};
\node[vertex, brown](x_3_3) at (0.0, -7.0) {};
\node[vertex, brown](y_3_2) at (-1.5, -7.0) {};
\node[vertex, blue](x_3_4) at (0.0, -7.5) {};
\node[vertex, blue](y_3_3) at (-1.5, -7.5) {};
\node[vertex, cyan](x_3_5) at (0.0, -8.0) {};
\node[vertex, cyan](y_3_4) at (-1.5, -8.0) {};
\node[vertex, green](x_4_1) at (0.0, -9.0) {};
\node[vertex, brown](x_4_2) at (0.0, -9.5) {};
\node[vertex, blue](x_4_3) at (0.0, -10.0) {};
\node[vertex, cyan](x_4_4) at (0.0, -10.5) {};
\node[vertex, red](x_4_5) at (0.0, -11.0) {};
\draw [p1, red] (x_1_1) -- (y_1_1);
\draw [p1, ] (y_1_1) -- (x_1_2);
\draw [p1, cyan] (x_1_2) -- (z_1_4);
\draw [p1, ] (z_1_4) -- (x_1_3);
\draw [p1, brown] (x_1_3) -- (z_1_5);
\draw [p1, ] (z_1_5) -- (x_1_4);
\draw [p1, green] (x_1_4) -- (y_1_2);
\draw [p1, ] (y_1_2) -- (x_1_5);
\draw [p1, blue] (x_1_5) -- (y_1_3);
\draw [p1, ] (y_1_3) -- (x_2_1);
\draw [p1, red] (x_2_1) -- (y_2_1);
\draw [p1, ] (y_2_1) -- (x_2_2);
\draw [p1, green] (x_2_2) -- (y_2_2);
\draw [p1, ] (y_2_2) -- (x_2_3);
\draw [p1, blue] (x_2_3) -- (z_2_4);
\draw [p1, ] (z_2_4) -- (x_2_4);
\draw [p1, cyan] (x_2_4) -- (z_2_5);
\draw [p1, ] (z_2_5) -- (x_2_5);
\draw [p1, brown] (x_2_5) -- (y_2_3);
\draw [p1, ] (y_2_3) -- (x_3_1);
\draw [p1, red] (x_3_1) -- (z_3_5);
\draw [p1, ] (z_3_5) -- (x_3_2);
\draw [p1, green] (x_3_2) -- (y_3_1);
\draw [p1, ] (y_3_1) -- (x_3_3);
\draw [p1, brown] (x_3_3) -- (y_3_2);
\draw [p1, ] (y_3_2) -- (x_3_4);
\draw [p1, blue] (x_3_4) -- (y_3_3);
\draw [p1, ] (y_3_3) -- (x_3_5);
\draw [p1, cyan] (x_3_5) -- (y_3_4);
\draw [p1, ] (y_3_4) -- (x_4_1);
\draw [p1, ] (x_4_1) -- (x_4_1);
\draw [p1, ] (x_4_1) -- (x_4_2);
\draw [p1, ] (x_4_2) -- (x_4_3);
\draw [p1, ] (x_4_3) -- (x_4_4);
\draw [p1, ] (x_4_4) -- (x_4_5);
\draw [p2, ] (x_1_1) -- (x_1_2);
\draw [p2, ] (x_1_2) -- (x_1_3);
\draw [p2, ] (x_1_3) -- (x_1_4);
\draw [p2, ] (x_1_4) -- (x_1_5);
\draw [p2, ] (x_1_5) -- (x_2_1);
\draw [p2, red] (x_2_1) -- (y_1_1);
\draw [p2, ] (y_1_1) -- (x_2_2);
\draw [p2, green] (x_2_2) -- (y_1_2);
\draw [p2, ] (y_1_2) -- (x_2_3);
\draw [p2, blue] (x_2_3) -- (y_1_3);
\draw [p2, ] (y_1_3) -- (x_2_4);
\draw [p2, cyan] (x_2_4) -- (z_1_4);
\draw [p2, ] (z_1_4) -- (x_2_5);
\draw [p2, brown] (x_2_5) -- (z_1_5);
\draw [p2, ] (z_1_5) -- (x_3_1);
\draw [p2, red] (x_3_1) -- (y_2_1);
\draw [p2, ] (y_2_1) -- (x_3_2);
\draw [p2, green] (x_3_2) -- (y_2_2);
\draw [p2, ] (y_2_2) -- (x_3_3);
\draw [p2, brown] (x_3_3) -- (y_2_3);
\draw [p2, ] (y_2_3) -- (x_3_4);
\draw [p2, blue] (x_3_4) -- (z_2_4);
\draw [p2, ] (z_2_4) -- (x_3_5);
\draw [p2, cyan] (x_3_5) -- (z_2_5);
\draw [p2, ] (z_2_5) -- (x_4_1);
\draw [p2, green] (x_4_1) -- (y_3_1);
\draw [p2, ] (y_3_1) -- (x_4_2);
\draw [p2, brown] (x_4_2) -- (y_3_2);
\draw [p2, ] (y_3_2) -- (x_4_3);
\draw [p2, blue] (x_4_3) -- (y_3_3);
\draw [p2, ] (y_3_3) -- (x_4_4);
\draw [p2, cyan] (x_4_4) -- (y_3_4);
\draw [p2, ] (y_3_4) -- (x_4_5);
\draw [p2, red] (x_4_5) -- (z_3_5);
	\end{tikzpicture}
		\caption{Construction of $G$ for $\pi = 43521$. We have $\sigma_1 = 14523$, $\sigma_2 = 12534$, $\sigma_3 = 23451$, $\sigma^{-1}_1 = 14523$, $\sigma^{-1}_2 = 12453$, $\sigma^{-1}_3 = 51234$, and $\pi_1 = 12345$, $\pi_2 = 14523$, $\pi_3 = 14352$, $\pi_4 = 43521$. (All permutations are of length $5$, we omitted commas and parentheses.)
		(\emph{left}) Vertices of $G$ with labels.
		(\emph{middle}) Illustration of the functions $s_x$ (dashed) and $s_c$ (colored by connected component).
		(\emph{right}) The actual graph $G$. Dashed edges belong to $P_1$, solid edges to $P_2$.}
	\label{fig:bad_3track}
\end{figure}

%%
%% Bibliography
%%

%% Please use bibtex, 
\newpage

\subsection{Proof of Lemma~\ref{lemhard} \label{appf}}

\begin{proof}

Let $G,H,\phi$ be an instance of PSI, as described above, with $n = |V(G)|$, $m = |E(G)|$ and $k=|V(H)|=|E(H)|$. Let $V(H) = \{v_1, \dots, v_k\}$. Let $V_1, \dots, V_k$ be a partitioning of $V(G)$, according to $\phi$, such that $u \in V_i$ if and only if $\phi(u) = v_i$. Consider a canonical ordering $(u_1, \dots, u_n)$ of the vertices of $G$, in which vertices in $V_i$ appear before vertices in $V_j$ for all $i<j$. Assume a preprocessing of $G$, where all edges with endpoints in the same class $V_i$ are deleted, for all $i\in [k]$, as these cannot be part of a subgraph of the required form. 

A PPPM instance is created, consisting of a text permutation $\tau$ of length $O(n^2)$, a pattern $\pi$ of length $O(k)$, and a coloring of $\tau$ with $|\pi|$ colors. Both $\tau$ and $\pi$ are based on a \emph{tilted grid} permutation. 
Intuitively, $\tau$ and $\pi$ represent the adjacency matrices of $G$ and $H$, as well as the coloring $\phi$ of $V(G)$, with minor additional gadgets that enforce that valid embeddings of $\pi$ into $\tau$ correspond exactly to selections of valid $H$-isomorphic subgraphs in $G$.

\medskip

We first describe the construction of $\pi$. Consider (conceptually) a grid of size $(k+1) \times (k+1)$, with cells indexed from $(0,0)$ (top left) to $(k,k)$ (bottom right).
In the first row, excluding cell $(0,0)$, place points forming the permutation $(2,1,4,3,6,5, \dots, 2k, 2k-1)$, with two consecutive points in each cell, left to right. 
In the first column, excluding cell $(0,0)$, place points forming the permutation $(2k-1, 2k, 2k-3, 2k-2, 2k-5, 2k-4, \dots, 1, 2)$, with two consecutive points in each cell, top to bottom.
Place one point in cell $(0,0)$ below all other points in its row, and to the left of all other points in its column. Place one point in each grid cell $(i,j)$, for $i,j \in [k]$ such that $i=j$ or $\{v_i,v_j\} \in E(H)$. The placement of a point in cell $(i,j)$ is such that it falls horizontally between the two points in cell $(i,0)$ and vertically between the two points in cell $(0,j)$. Furthermore the points in cells $(i,j)$ for $i,j \in [k]$ are in each row increasing by $y$-coordinate left-to-right, and are in each column increasing by $x$-coordinate top-to-bottom. Let $\pi$ denote the permutation that has the same ordering as the constructed point set. It is easy to verify that $\pi$ is of length $7k+1$. See Figure~\ref{fighard} for an example.

\medskip

We next describe the construction of $\tau$. Consider again a grid of size $(k+1) \times (k+1)$, with cells indexed from $(0,0)$ (top left) to $(k,k)$ (bottom right).
In the first row, excluding cell $(0,0)$, place points forming the permutation $(2,1,4,3,6,5, \dots, 2n, 2n-1)$, with $2|V_i|$ consecutive points in cell $(i,0)$, left to right. 
In the first column, excluding cell $(0,0)$, place points forming the permutation $(2n-1, 2n, 2n-3, 2n-2, 2n-5, 2n-4, \dots, 1, 2)$, with $2|V_i|$ consecutive points in cell $(0,i)$, top to bottom.
Place one point in cell $(0,0)$ below all other points in its row, and to the left of all other points in its column. Place additional points, horizontally between the $(2i)$-th and $(2i+1)$-th points of the first row and vertically between the $(2j)$-th and $(2j+1)$-th points of the first column, for $i,j \in [n]$ such that $i=j$, or $\{u_i,u_j\} \in E(G)$. The placement of the points in cells $(\ell,h)$ for $\ell,h \in [k]$ is such that they are in each row increasing by $y$-coordinate left-to-right, and they are in each column increasing by $x$-coordinate top-to-bottom. Let $\tau$ denote the permutation that has the same ordering as the constructed point set. It is easy to verify that $\tau$ is of length $5n+2m+1 \in O(n^2)$. %See Figure~\ref{fighard} for an example.

\begin{figure*}
  \centering
  \includegraphics[width=14cm]{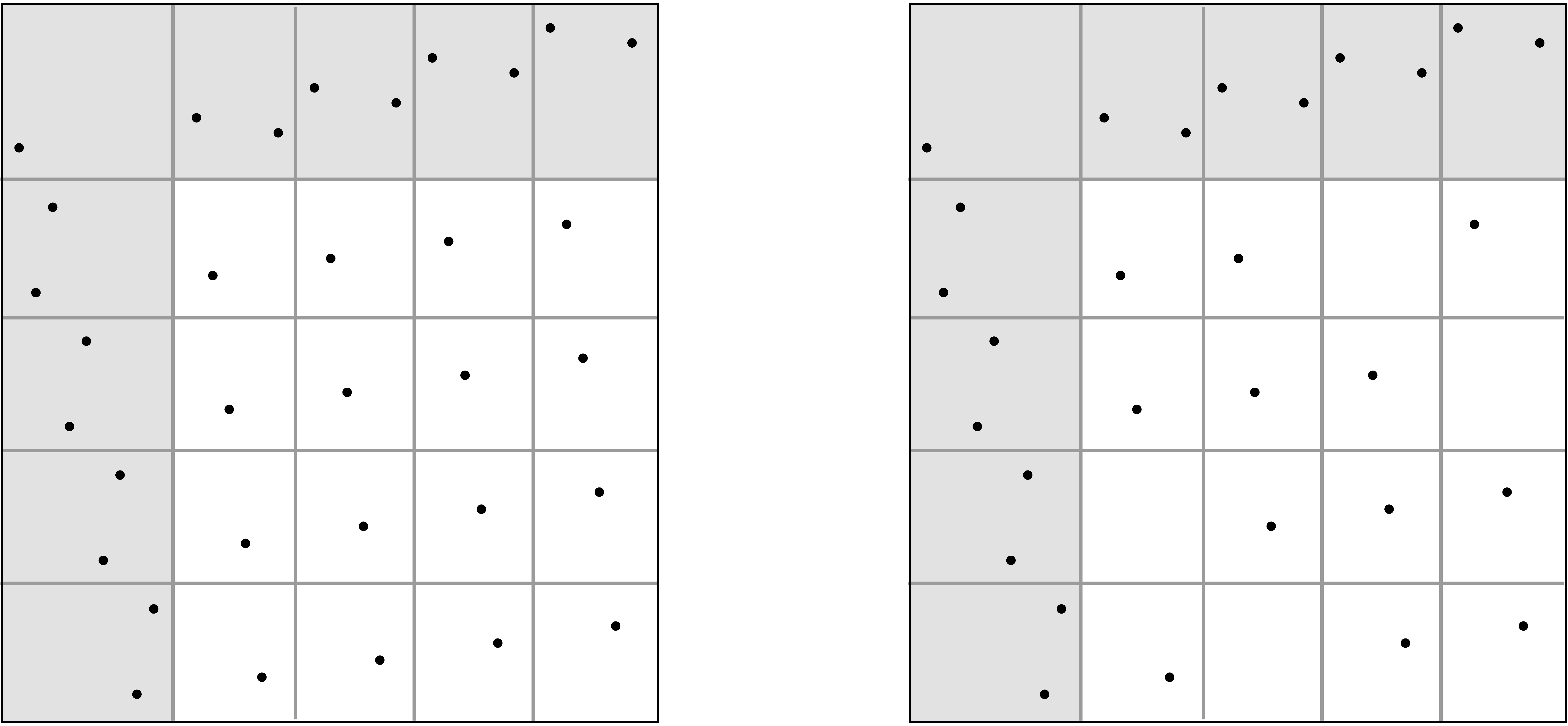}
  \caption{(\emph{left}) Pattern corresponding to clique $K_4$. (\emph{right}) Pattern corresponding to cycle $C_4$. First row and column highlighted for clarity.\label{fighard}}
\end{figure*}

\medskip

It remains to specify the coloring of $\tau$ with $k$ colors. Color all points in cell $(i,j)$ of the text-grid, for $i,j \in [k]$, with the index of the unique point in cell $(i,j)$ of the pattern-grid. Cells $(i,0)$ and $(0,i)$ of the pattern-grid, for $i \in [k]$ have two points each. Denote the indices of these as $a_{(i,0)}$ and $b_{(i,0)}$, resp.\ $a_{(0,i)}$ and $b_{(0,i)}$, left-to-right, resp.\ top-to-bottom. Color points in cell $(i,0)$ of the text-grid, for $i \in [k]$, with $a_{(i,0)}$ and $b_{(i,0)}$, alternatingly left-to-right. Color points in cell $(0,i)$ of the text-grid, for $i \in [k]$, with $a_{(0,i)}$ and $b_{(0,i)}$, alternatingly top-to-bottom. Finally, color the unique point in the text-cell $(0,0)$ with the index of the unique point in the pattern-cell $(0,0)$.

This concludes the construction of the PPPM instance. 

\medskip

Suppose there is a subgraph of $G$ that is isomorphic to $H$, with an embedding of the vertices that respects the coloring of $V(G)$. Let $u_{i_1}, \dots, u_{i_k} \in V(G)$ be the vertices matched, respectively, to $v_1, \dots, v_k \in V(H)$. Observe that $u_{i_j} \in V_j$ for all $j \in [k]$. We argue that there is a valid embedding of $\pi$ into $\tau$ that respects the coloring of $\tau$. 
Indeed, match the $(2j)$th and $(2j+1)$th points in the first row of the pattern-cell to the $(2i_j)$th and $(2i_j + 1)$th points in the first row of the text-cell. Assuming the canonical ordering of vertices, the coloring of these points is in accordance with the matching. Similarly, match the $(2j)$th and $(2j+1)$th points in the first column of the pattern-cell to the $(2i_j)$th and $(2i_j + 1)$th points in the first column of the text-cell. Assuming the canonical ordering of vertices, the coloring of these points is in accordance with the matching. If there is a point in the pattern-cell $(\ell,h)$ for $\ell,h \in [k]$, then $\{v_\ell,v_h\} \in E(H)$, and then, by assumption, $\{u_{i_\ell},u_{i_h}\} \in E(G)$. Consequently, there is a unique point in the text-cell $(\ell,h)$ that has the correct color, and that is horizontally between the matched points in cell $(\ell,0)$ and vertically between the matched points in cell $(0,h)$. 
Finally, match the unique point in the pattern-cell $(0,0)$ to the unique point in the text-cell $(0,0)$. The fact that the text-points used in the embedding have the correct relative positions is easy to verify.

\medskip

Conversely, suppose there is a valid, color-respecting embedding of $\pi$ into $\tau$. Then, pairs of points in the pattern-cells $(1,0),\dots,(k,0)$ and $(0,1),\dots,(0,k)$ must be matched to consecutive points in the corresponding text-cells of colors $a_{(*,*)}$ and $b_{(*,*)}$  (since only such pairs are in the correct relative position). These pairs of points determine a selection of vertices $u_{i_1}, \dots, u_{i_k}$ from the classes $V_1, \dots, V_k$, respectively. The existence of a point in a pattern-cell $(\ell,h)$ for $\ell,h \in [k]$ means that $\{v_\ell, v_h\} \in E(H)$. Such a point can only be matched to a point in the text-cell $(\ell,h)$, and since the embedding has the valid ordering, this point must be bracketed by the pairs of points in the first row and first column that correspond to the vertices $u_{i_\ell}$, resp.\ $u_{i_h}$. The existence of such a point in the text-pattern means that indeed, the corresponding edge $\{u_{i_\ell},u_{i_h}\}$ exists in $G$. 
\end{proof} 

\newpage

\bibliography{submission}

\end{document}